\documentclass[twocolumn]{aa}
\usepackage{graphicx}
\usepackage{natbib,lscape}
\begin{document}
   \title{XMM-Newton observations of High Mass X-ray Binaries in the SMC
\thanks{XMM-Newton is an ESA Science Mission
with instruments and contributions directly funded by ESA Member
states and the USA (NASA).}}         
   \titlerunning{XMM-Newton observations of HMXBs in the SMC}

   \author{M. Sasaki\thanks{\emph{Present address:}
               Harvard-Smithsonian Center for Astrophysics, 60 Garden Street,
               Cambridge, MA 02138, USA}
	\and W.\ Pietsch 
	\and F.\ Haberl }

   \authorrunning{M. Sasaki et al.}

   \offprints{M. Sasaki, \email{msasaki@cfa.harvard.edu}}
 
   \institute{Max-Planck-Institut f\"ur extraterrestrische Physik,
               Giessenbachstra{\ss}e, Postfach 1312, 85741 Garching, Germany}

   \date{Received December, 11, 2002; accepted March, 17, 2003}

   \abstract{
Based on XMM-Newton EPIC data of four pointings towards the  
Small Magellanic Cloud (SMC), results on timing and spectral analyses of 16 
known high mass X-ray binaries (HMXBs) and HMXB candidates in the SMC are 
presented. We confirm the pulse periods of four sources which were known to 
show pulsations. In addition, two new X-ray pulsars are discovered:
XMMU\,J005605.2--722200 with $P_{\rm pulse} = 140.1\pm0.3$~s and 
RX\,J0057.8--7207 with $P_{\rm pulse} = 152.34\pm0.05$~s. 
Due to the low Galactic foreground absorption, X-ray binary systems in the 
Magellanic Clouds are well suited for studies of 
the soft component in their X-ray spectrum. 
Spectral analysis reveals soft emission besides a power law component in the 
spectra of three sources. The existence of 
emission lines in at least one of them 
corroborates
the thermal nature of this emission with temperatures of 0.2 -- 0.3~keV 
and heavy element abundances lower than solar.
For the HMXB SMC\,X-2 which was in a low luminosity state, 
we determine a flux upper limit of 
$1.5\times 10^{-14}$~erg~cm$^{-2}$~s$^{-1}$ (0.3 -- 10.0~keV).
Furthermore, two new sources (XMMU\,J005735.7--721932 and 
XMMU\,J010030.2--722035) with hard spectrum and emission line objects 
as likely optical counterparts are proposed as new X-ray binary candidates. 
   }

   \maketitle
%
%________________________________________________________________

\section{Introduction}

After the discovery of X-ray emission from the Magellanic Clouds (MCs)
in 1970 \citep{1971ApJ...168L...7P}, surveying observations of each
MC were performed by different X-ray observatories. As for the Small
Magellanic Cloud (SMC), source catalogues were created from observations with
Einstein \citep{1981ApJ...243..736S, 
1987ApJ...317..152B,1992ApJS...78..391W}, ROSAT 
\citep{1999A&AS..136...81K,2000A&AS..142...41H,2000A&AS..147...75S}, 
and ASCA \citep{2000ApJS..128..491Y}.

The analysis of these X-ray sources has shown, that a large number of 
X-ray bright objects belongs to the class of  
X-ray binaries (XRBs) in which a neutron star or a black hole forms a binary 
system with a companion star. In these systems, mass is accreted from the
donor star onto the compact object. X-ray binaries can be 
divided into low mass X-ray binaries and high mass 
(or massive) X-ray binaries (HMXBs), depending on the mass of the
companion star. Therefore, the identification of optical 
counterparts of the X-ray sources is crucial for the understanding
of the nature of these sources. Furthermore, HMXBs form two subgroups
with either an OB 
supergiant or a Be star as donor. A detailed catalogue of HMXBs was compiled by
\citet{2000A&AS..147...25L}. \citet{2002A&A...385..517N} performed high
resolution spectroscopy of optical counterparts of HMXBs in the Large 
Magellanic Cloud
(LMC) and studied the population of HMXBs. In the Milky Way or in
the LMC, the fraction of Be/X-ray binary systems (Be/XRB) is
60 -- 70\% of all HMXBs, whereas more than 90\% of the HMXBs in the SMC 
turned out to be Be systems
\citep[][and references therein]{2000A&A...359..573H}.

Since pulsed X-ray emission can be observed from neutron star HMXBs, these
sources are also called X-ray binary pulsars. Based on ASCA, RXTE, ROSAT and 
Beppo SAX observations, more than 20 X-ray binary pulsars have been 
discovered in the SMC so far 
\citep[][and references therein]{2000A&A...359..573H,
2000ApJS..128..491Y}. Moreover, in one of the first observations
of XMM-Newton \citep{2001A&A...365L...1J}, pulsed emission from 
another HMXB was found, which was identified 
with a Be star \citep{2001A&A...369L..29S}.

In order to improve our understanding of the X-ray source population in
the SMC, we proposed and 
analysed pointed observations of the SMC by XMM-Newton and 
performed spectral and temporal studies of detected sources. 
In this paper, we focus on the class of HMXBs and present the results on 
each HMXB and candidate in the observed fields. 

\section{Data}

\begin{table*}[t]
\caption[]{\label{obstab} XMM-Newton data used for the analysis.}
\begin{tabular}{clcccccll}
\hline\noalign{\smallskip}
\multicolumn{1}{c}{1} & \multicolumn{1}{c}{2} & \multicolumn{2}{c}{3} & \multicolumn{1}{c}{4} &
\multicolumn{1}{c}{5} & \multicolumn{1}{c}{6} & \multicolumn{1}{c}{7} & \multicolumn{1}{c}{8} \\
\hline\noalign{\smallskip}
\multicolumn{1}{c}{Rev.} & \multicolumn{1}{c}{Obs.\ ID} & \multicolumn{2}{c}{Pointing direction} & \multicolumn{1}{c}{Inst.} & \multicolumn{1}{c}{Mode} & \multicolumn{1}{c}{Filter} & \multicolumn{1}{c}{Start time (UT)} & \multicolumn{1}{c}{End time (UT)} \\
\multicolumn{1}{c}{} & \multicolumn{1}{c}{} & \multicolumn{1}{c}{RA} & \multicolumn{1}{c}{Dec} & \multicolumn{1}{c}{(EPIC)} & \multicolumn{1}{c}{} & \multicolumn{1}{c}{} & \multicolumn{1}{c}{} & \multicolumn{1}{c}{} \\
\multicolumn{1}{c}{} & \multicolumn{1}{c}{} & \multicolumn{2}{c}{(J2000.0)} & \multicolumn{1}{c}{} & \multicolumn{1}{c}{} & \multicolumn{1}{c}{} & \multicolumn{1}{c}{} & \multicolumn{1}{c}{} \\
\noalign{\smallskip}\hline\noalign{\smallskip}                       
157 & 01100002 & 00 59 46.6 & --72 09 30 & PN & Full  & Medium & 2000/10/17 16:16:36 & 2000/10/17 20:41:09 \\
    &            &            &            & M1 & Full  & Medium  & 2000/10/17 15:10:44 & 2000/10/17 20:39:43 \\
    &            &            &            & M2 & Full  & Medium  & 2000/10/17 15:10:35 & 2000/10/17 20:39:42 \\
247 & 01357206 & 01 03 29.0 & --72 02 33 & PN & Full  & Thin1 & 2001/04/15 01:20:28 & 2001/04/15 05:50:27 \\
    &            &            &            & M1 & PW3 & Thin1 & 2001/04/14 20:47:25 & 2001/04/15 05:55:45 \\
    &            &            &            & M2 & PW3 & Thin1 & 2001/04/14 20:47:25 & 2001/04/15 05:55:45 \\
340 & 00842008 & 00 54 54.3 & --73 40 12 & PN & Full  & Thin1 & 2001/10/17 10:46:50 & 2001/10/17 15:55:11 \\
    &            &            &            & M1 & Full  & Medium  & 2001/10/17 10:07:40 & 2001/10/17 15:59:35 \\
    &            &            &            & M2 & Full  & Medium  & 2001/10/17 10:07:40 & 2001/10/17 15:59:36 \\
422 & 00842001 & 00 56 24.4 & --72 21 33 & PN & Full  & Thin1 & 2002/03/30 14:21:45 & 2002/03/30 19:39:38 \\
    &            &            &            & M1 & Full  & Medium  & 2002/03/30 13:48:28 & 2002/03/30 19:44:36 \\
    &            &            &            & M2 & Full  & Medium  & 2002/03/30 13:48:29 & 2002/03/30 19:44:54 \\
\noalign{\smallskip}\hline
\end{tabular}

\vspace{1.5mm}
Notes to column No 5:
PW3: Partial Window 3.

\vspace{.5mm}
Notes to column No 6:
M1: MOS1, M2: MOS2. 
\end{table*}

\begin{figure}
\centering
\includegraphics[width=8.5cm,bb=28 55 545 480]{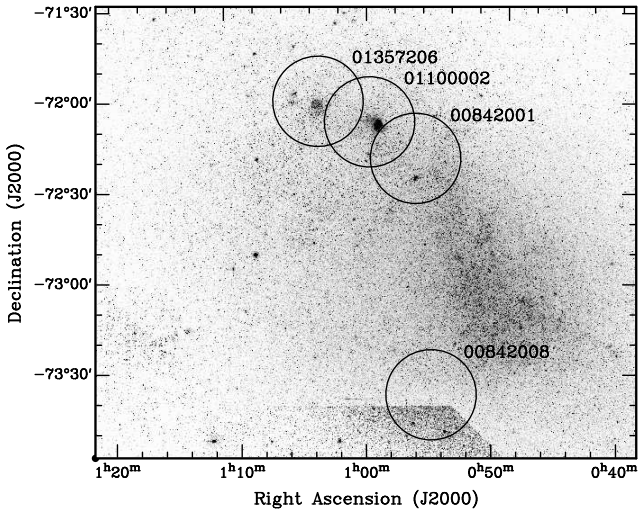}
\caption{\label{pointings}
DSS image of the SMC and the position of the EPIC field of view of the 
XMM-Newton observations listed in Table \ref{obstab}. The discontinuity seen 
in the DSS image is an artifact.}
\end{figure}                                  

For AO-1 of XMM-Newton, we proposed observations of eight fields in the SMC
in order to study the X-ray binary population (PI: W.P.). Two 
observations of this proposal were performed. 
During the first observation (ID 00842008), the telescope pointed towards  
the HMXB SMC\,X-2 in the south of the main galaxy.
In order to perform source detection and analysis in the whole field of 
view, we used data from the European Photon Imaging Cameras EPIC 
PN \citep{2001A&A...365L..18S}, EPIC MOS1, and EPIC MOS2 
\citep{2001A&A...365L..27T}.
The observation was performed with all the EPIC cameras 
in full frame mode. For EPIC PN the thin filter was
used, whereas for the EPIC MOS cameras medium filters were chosen.       
The next observation, ID 00842001, covered a region in the north
of the SMC. 
The CCD read out modes and the filters of the EPIC cameras 
were the same as in the first 
observation.

Moreover, we searched the XMM-Newton Science Data Archive for public data of 
the SMC, suitable for our purposes. We found two data sets (ID 01100002 and 
01357206) of fields in the north of the SMC, slightly 
overlapping with each other as well as with the pointing ID 00842001. 
The details of the observations are summarised in Table \ref{obstab}.

\subsection{Source detection}\label{soudet}

\begin{table*}[t]
\caption[]{\label{xbtab}
Detected HMXBs and candidates in the fields of the SMC observed by XMM-Newton.
}
\begin{tabular}{lrllccrc}
\hline\noalign{\smallskip}
\multicolumn{1}{c}{1} & \multicolumn{1}{c}{2} & \multicolumn{1}{c}{3} & \multicolumn{1}{c}{4} &
\multicolumn{1}{c}{5} & \multicolumn{1}{c}{6} & \multicolumn{1}{c}{7} & \multicolumn{1}{c}{8} \\
\hline\noalign{\smallskip}
\multicolumn{1}{c}{No} & \multicolumn{1}{c}{Obs.\ ID} & \multicolumn{1}{c}{RA} &
\multicolumn{1}{c}{Dec} & \multicolumn{1}{c}{Pos.\ err.} & \multicolumn{1}{c}{Count rate} & \multicolumn{1}{c}{Det. ML} & 
\multicolumn{1}{c}{Flux} \\
\multicolumn{1}{c}{} & \multicolumn{1}{c}{} & \multicolumn{2}{c}{(J2000.0)} &
\multicolumn{1}{c}{[\arcsec]} & \multicolumn{1}{c}{[s$^{-1}$]} & \multicolumn{1}{c}{} & 
\multicolumn{1}{c}{[erg~cm$^{-2}$~s$^{-1}$]} \\
\noalign{\smallskip}\hline\noalign{\smallskip}                   
01 & 00842008\,M1 & 00 51 56.05 & --73 41 51.4 & 2.1 & $6.33 \times 10^{-3} \pm 1.04 \times 10^{-3}$ & 84.2 &$\sim1 \times 10^{-13}$\\
\noalign{\smallskip}\hline\noalign{\smallskip}       
02 & 00842008\,PN & 00 54 33.4  & --73 41 04$^{\diamondsuit}$& -- & $<2.33 \times 10^{-03\,\dagger}$ & 3.4 & $<1.5 \times 10^{-14}$\\
\noalign{\smallskip}\hline\noalign{\smallskip}      
03 & 00842001\,PN & 00 54 56.02 & --72 26 48.6 & 1.0 & $3.30 \times 10^{-2} \pm 3.44 \times 10^{-3}$ & 267.1 & $5.3 \times 10^{-14}$ \\
\noalign{\smallskip}\hline\noalign{\smallskip}     
04 & 00842001\,PN & 00 56 05.24 & --72 22 00.9 & 2.0 & $7.46 \times 10^{-3} \pm 1.32 \times 10^{-3}$ &  76.0 & $2.4 \times 10^{-14}$ \\
\noalign{\smallskip}\hline\noalign{\smallskip}    
05 & 00842001\,PN & 00 57 19.58 & --72 25 35.1 & 0.5 & $8.79 \times 10^{-2} \pm 4.16 \times 10^{-3}$ & 225.6 & $1.4 \times 10^{-13}$ \\
\noalign{\smallskip}\hline\noalign{\smallskip}   
06 & 00842001\,PN & 00 57 35.71 & --72 19 32.6 & 0.9 & $3.07 \times 10^{-2} \pm 2.46 \times 10^{-3}$ & 478.5 & $9.5 \times 10^{-14}$ \\
06a& 01100002\,PN & 00 57 35.56 & --72 19 36.8 & 2.9 & $2.53 \times 10^{-2} \pm 6.94 \times 10^{-3}$ &  55.9 & $1.9 \times 10^{-14}$ \\
\noalign{\smallskip}\hline\noalign{\smallskip}  
07 & 01100002\,PN & 00 57 50.22 & --72 02 37.0 & 0.7 & $1.32 \times 10^{-1} \pm 8.75 \times 10^{-3}$ &1147.0 & $2.3 \times 10^{-13}$ \\
\noalign{\smallskip}\hline\noalign{\smallskip} 
08 & 01100002\,PN & 00 57 50.80 & --72 07 58.7 & 0.5 & $1.79 \times 10^{-1} \pm 8.11 \times 10^{-3}$ &2798.1 & $5.8 \times 10^{-13}$ \\
\noalign{\smallskip}\hline\noalign{\smallskip}            
09 & 00842001\,PN & 00 58 11.68 & --72 30 50.4 & 1.7 & $5.37 \times 10^{-2} \pm 8.49 \times 10^{-3}$ & 120.2 & $\sim3 \times 10^{-13}$ \\
\noalign{\smallskip}\hline\noalign{\smallskip}           
10 & 01100002\,M1 & 00 59 21.01 & --72 23 18.4 & 0.8 & $5.22 \times 10^{-2} \pm 3.22 \times 10^{-3}$ &1269.6 & $2.3 \times 10^{-13}$ \\
10a& 00842001\,PN & 00 59 21.10 & --72 23 15.8 & 0.9 & $1.05 \times 10^{-1} \pm 7.23 \times 10^{-3}$ &828.9  & $1.2 \times 10^{-13}$ \\
\noalign{\smallskip}\hline\noalign{\smallskip}          
11 & 01100002\,PN & 01 00 16.18 & --72 04 45.8 & 1.2 & $2.42 \times 10^{-2} \pm 2.86 \times 10^{-3}$ & 237.5 & $4.2 \times 10^{-14}$ \\
\noalign{\smallskip}\hline\noalign{\smallskip}         
12 & 01100002\,PN & 01 00 30.23 & --72 20 35.1 & 3.2 & $9.81 \times 10^{-3} \pm 3.00 \times 10^{-3}$ &  25.8 &$\sim6 \times 10^{-14}$\\
\noalign{\smallskip}\hline\noalign{\smallskip}         
13 & 01357206\,PN & 01 01 02.98 & --72 06 59.5 & 2.9 & $1.77 \times 10^{-2} \pm 4.16 \times 10^{-3}$ &  38.2 & $2.6 \times 10^{-14}$ \\
13a& 01100002\,PN & 01 01 03.87 & --72 07 02.4 & 3.8 & $9.16 \times 10^{-3} \pm 2.68 \times 10^{-3}$ &  23.0 & $3.2 \times 10^{-14}$ \\
\noalign{\smallskip}\hline\noalign{\smallskip}        
14 & 01100002\,PN & 01 01 20.82 & --72 11 21.1 & 0.6 & $1.40 \times 10^{-1} \pm 7.47 \times 10^{-3}$ &1689.6 & $3.1 \times 10^{-13}$ \\
14a& 01357206\,PN & 01 01 20.87 & --72 11 16.8 & 1.0 & $8.83 \times 10^{-2} \pm 8.74 \times 10^{-3}$ & 368.7 & $8.5 \times 10^{-14}$ \\
\noalign{\smallskip}\hline\noalign{\smallskip}       
15 & 01357206\,PN & 01 01 37.56 & --72 04 18.7 & 1.0 & $5.01 \times 10^{-2} \pm 4.26 \times 10^{-3}$ & 436.3 & $5.7 \times 10^{-14}$ \\
15a& 01100002\,PN & 01 01 37.77 & --72 04 22.1 & 1.2 & $3.55 \times 10^{-2} \pm 4.36 \times 10^{-3}$ & 222.7 & $4.6 \times 10^{-14}$ \\
\noalign{\smallskip}\hline\noalign{\smallskip}      
16 & 01357206\,PN & 01 03 14.11 & --72 09 14.2 & 0.5 & $1.34 \times 10^{-1} \pm 6.38 \times 10^{-3}$ &2049.8 & $3.5 \times 10^{-13}$ \\
\noalign{\smallskip}\hline\noalign{\smallskip} 
17 & 01357206\,PN & 01 03 37.57 & --72 01 33.2 & 0.2 & $5.09 \times 10^{-1} \pm 9.62 \times 10^{-3}$ &17172.3& $2.6 \times 10^{-12}$ \\
\noalign{\smallskip}\hline\noalign{\smallskip}     
18 & 01357206\,PN & 01 05 55.38 & --72 03 47.9 & 1.6 & $3.05 \times 10^{-2} \pm 3.79 \times 10^{-3}$ & 119.5 & $3.9 \times 10^{-14}$ \\
\noalign{\smallskip}
\hline
\end{tabular}       

\vspace{1.5mm}
Notes to column No 1:
Observation ID and the instrument with which the detection with the highest 
likelihood was achieved.

\vspace{.5mm}
Notes to columns No 3 -- 5:
Position of the XMM-Newton detections with 1~$\sigma$ statistical positional 
error, except for No 02, for which the position of the optical 
counterpart$^{\diamondsuit}$
is given.

\vspace{.5mm}
Notes to column No 6:
Count rate for EPIC PN detection (except for entries No 01 and 10 which are 
EPIC MOS1 detections, and $^{\dagger}$ 3~$\sigma$ upper limit for No 02).

\vspace{.5mm}
Notes to column No 7:
Maximum likelihood (ML) of detection for the total band (0.3 -- 10.0~keV). 

\vspace{.5mm}
Notes to column No 8:
Flux (0.3 -- 10.0~keV) calculated from the best fit model spectrum, setting
the Galactic foreground absorption to zero. See Sect.\,\ref{timspec} 
for used models. In order to obtain the luminosity, multiply by 
$4.3 \times 10^{47}$~cm$^{2}$.
\end{table*}

\addtocounter{table}{-1}  
\begin{table*}[t]
\caption[]{
Continued.}     
\begin{tabular}{lccccccc}
\hline\noalign{\smallskip}
\multicolumn{1}{c}{1} & \multicolumn{1}{c}{9} & \multicolumn{1}{c}{10} & \multicolumn{1}{c}{11} &
\multicolumn{1}{c}{12} & \multicolumn{1}{c}{13} & \multicolumn{1}{c}{14} & \multicolumn{1}{c}{15} \\
\hline\noalign{\smallskip}
\multicolumn{1}{c}{No} & \multicolumn{1}{c}{HR1} & \multicolumn{1}{c}{HR2} & \multicolumn{1}{c}{HR3} & 
\multicolumn{1}{c}{P} & \multicolumn{1}{c}{$d_{\rm O}$} & \multicolumn{1}{c}{OGLE Name} & \multicolumn{1}{c}{[MA93]}\\
\multicolumn{1}{c}{} & \multicolumn{1}{c}{} & \multicolumn{1}{c}{} &
\multicolumn{1}{c}{} & \multicolumn{1}{c}{[s]} & 
\multicolumn{1}{c}{[\arcsec]} &\multicolumn{1}{c}{} & \multicolumn{1}{c}{} \\
\noalign{\smallskip}\hline\noalign{\smallskip}                   
01 &+0.15$\pm$0.20 &+0.09$\pm$0.18 &--1.00$\pm$0.26 &  --                              & --  & --                &  --  \\
\noalign{\smallskip}\hline\noalign{\smallskip}                                                                 
02 & --            & --            & --             &  too weak                        & --  & --                & --   \\
\noalign{\smallskip}\hline\noalign{\smallskip}                                                                
03 &+0.41$\pm$0.12 &--0.03$\pm$0.11 &--0.30$\pm$0.17&  59.00$\pm$0.02                  & 1.2 & 00545617--7226476 &  810 \\
\noalign{\smallskip}\hline\noalign{\smallskip}                                                               
04 &+1.00$\pm$0.23 &--0.15$\pm$0.17 &--0.13$\pm$0.25&  140.1$\pm$0.3*                  & --  & --                &  904 \\
\noalign{\smallskip}\hline\noalign{\smallskip}                                                              
05 &+0.49$\pm$0.05 &--0.29$\pm$0.05 &--0.36$\pm$0.08& --                               & 1.8 & 00571981--7225337 & --   \\
\noalign{\smallskip}\hline\noalign{\smallskip}                                                             
06 &+0.45$\pm$0.11 & +0.04$\pm$0.09 &--0.41$\pm$0.11&                                  & 1.9 &                   &      \\
06a&+0.80$\pm$0.31 & +0.10$\pm$0.26 &--0.54$\pm$0.43& \raisebox{1.5mm}[.5mm][.5mm]{--} & 3.5 & \raisebox{1.5mm}[.5mm][.5mm]{00573601--7219339}& \raisebox{1.5mm}[.5mm][.5mm]{1020} \\    
\noalign{\smallskip}\hline\noalign{\smallskip}                                                            
07 &+0.19$\pm$0.09 & +0.03$\pm$0.08 &--0.14$\pm$0.10&  281.1$\pm$0.2                   & --  & --                & 1036 \\
\noalign{\smallskip}\hline\noalign{\smallskip}                                                           
08 &+0.48$\pm$0.07 & +0.25$\pm$0.05 &--0.06$\pm$0.06& 152.34$\pm$0.05*                 & --  & --                & 1038 \\
\noalign{\smallskip}\hline\noalign{\smallskip}                                                          
09 &+1.00$\pm$0.33 & +0.44$\pm$0.21 & +0.51$\pm$0.13& --                               & 4.5 & 00581258--7230485 & --   \\
\noalign{\smallskip}\hline\noalign{\smallskip}                                                         
10 &+0.49$\pm$0.07 &--0.15$\pm$0.07 &--0.21$\pm$0.10&                                  & 1.2 &                   &      \\
10a&+0.27$\pm$0.09 &--0.05$\pm$0.09 & +0.12$\pm$0.09&  \raisebox{1.5mm}[.5mm][.5mm]{--}& 1.4 & \raisebox{1.5mm}[.5mm][.5mm]{00592103--7223171}& \raisebox{1.5mm}[.5mm][.5mm]{--} \\
\noalign{\smallskip}\hline\noalign{\smallskip}                                                        
11 &+0.18$\pm$0.14 &--0.19$\pm$0.14 &--0.46$\pm$0.21& --                               & --  &  --               & --   \\
\noalign{\smallskip}\hline\noalign{\smallskip}                                                       
12 &+1.00$\pm$0.13 &--0.37$\pm$0.29 & +0.01$\pm$0.52&  --                              & --  &  --               & 1208 \\         
\noalign{\smallskip}\hline\noalign{\smallskip}                                                      
13 &+0.37$\pm$0.29 &--0.30$\pm$0.25 & +0.06$\pm$0.41&                                  &     &                   &      \\
13a&--0.25$\pm$0.36 &--0.21$\pm$0.48 &+0.40$\pm$0.41& \raisebox{1.5mm}[.5mm][.5mm]{--} & \raisebox{1.5mm}[.5mm][.5mm]{--} & \raisebox{1.5mm}[.5mm][.5mm]{--}& \raisebox{1.5mm}[.5mm][.5mm]{1240} \\
\noalign{\smallskip}\hline\noalign{\smallskip}                                                     
14 &+0.14$\pm$0.08 & +0.21$\pm$0.07 & +0.06$\pm$0.07&  452.2$\pm$0.5                   & 2.6 &                   &      \\
14a&+0.26$\pm$0.12 & +0.06$\pm$0.12 &--0.10$\pm$0.15& too weak                         & 2.2 & \raisebox{1.5mm}[.5mm][.5mm]{01012064--7211187}& \raisebox{1.5mm}[.5mm][.5mm]{1257} \\
\noalign{\smallskip}\hline\noalign{\smallskip}                                                    
15 &+0.17$\pm$0.10 &--0.23$\pm$0.10 &--0.20$\pm$0.16&                                  &     &                   &      \\
15a&+0.17$\pm$0.13 &--0.36$\pm$0.12 &--0.26$\pm$0.30& \raisebox{1.5mm}[.5mm][.5mm]{--} &\raisebox{1.5mm}[.5mm][.5mm]{--} &\raisebox{1.5mm}[.5mm][.5mm]{--} &\raisebox{1.5mm}[.5mm][.5mm]{1277}  \\
\noalign{\smallskip}\hline\noalign{\smallskip}                                                   
16 &+0.16$\pm$0.07 & +0.11$\pm$0.06 & +0.00$\pm$0.06&  341.7$\pm$0.4                   & --  & --                & 1367 \\
\noalign{\smallskip}\hline\noalign{\smallskip}                                                  
17 &+0.28$\pm$0.03 & +0.18$\pm$0.02 & +0.07$\pm$0.02& --                               & --  & --                & 1393 \\
\noalign{\smallskip}\hline\noalign{\smallskip}                                                 
18 &--0.19$\pm$0.18 & +0.12$\pm$0.18 &+0.22$\pm$0.16& --                               & --  & --                & 1557 \\    
\noalign{\smallskip}
\hline
\end{tabular}       

\vspace{1.5mm}
Notes to columns No 9 -- 11:
Hardness ratios as defined in Eq.\,(\ref{hr123}). 

\vspace{.5mm}
Notes to column No 12:
Pulse periods from timing analysis. * New X-ray binary pulsar!

\vspace{.5mm}
Notes to column No 13:
Distance to OGLE object.

\vspace{.5mm}
Notes to column No 15:
Entry numbers in \citet{1993A&AS..102..451M}.
\end{table*}

\addtocounter{table}{-1}  
\begin{table*}[t]
\caption[]{
Continued.}     
\begin{tabular}{lccccll}
\hline\noalign{\smallskip}
\multicolumn{1}{c}{1} & 
\multicolumn{1}{c}{16} & \multicolumn{1}{c}{17} & \multicolumn{1}{c}{18} & 
\multicolumn{1}{c}{19} & \multicolumn{1}{c}{20} & \multicolumn{1}{c}{21} \\
\hline\noalign{\smallskip}
\multicolumn{1}{c}{No} & \multicolumn{1}{c}{$d_{\rm R}$} 
& \multicolumn{1}{c}{HRI} & \multicolumn{1}{c}{PSPC} & \multicolumn{1}{c}{[HS2000]}  
& \multicolumn{1}{c}{Source ID} & \multicolumn{1}{c}{Remarks} \\
\multicolumn{1}{c}{} & \multicolumn{1}{c}{[\arcsec]} & 
\multicolumn{1}{c}{} & \multicolumn{1}{c}{} & \multicolumn{1}{c}{} & 
\multicolumn{1}{c}{} & \multicolumn{1}{c}{} \\
\noalign{\smallskip}\hline\noalign{\smallskip}                   
01 & --  & --  & --  & -- & RX\,J0051.7--7341  & XRB? \\
\noalign{\smallskip}\hline\noalign{\smallskip}                       
02 &22.8 & --  & 547 & 28 & SMC\,X-2          & HMXB Be, P \\
\noalign{\smallskip}\hline\noalign{\smallskip}                       
03 & 4.5 & 058 & 241 & 31 & RX\,J0054.9--7226 & HMXB Be, P \\
\noalign{\smallskip}\hline\noalign{\smallskip}                       
04 & --  & --  & --  & 32 & XMMU\,J005605.2--722200 = 2E\,0054.4--7237?  & HMXB?, P \\
\noalign{\smallskip}\hline\noalign{\smallskip}                       
05 & 5.6 & --  & 234 & -- & 2E\,0055.6--7241, RX\,J0057.3--7225 & AGN, $z$ = 0.15 \\
\noalign{\smallskip}\hline\noalign{\smallskip}                       
06 &     &     &     &    &  &  \\
06a& \raisebox{1.5mm}[.5mm][.5mm]{--} & \raisebox{1.5mm}[.5mm][.5mm]{--} & \raisebox{1.5mm}[.5mm][.5mm]{--} & \raisebox{1.5mm}[.5mm][.5mm]{--} & \raisebox{1.5mm}[.5mm][.5mm]{XMMU\,J005735.7--721932 = [YIT2000]\,19?} & \raisebox{1.5mm}[.5mm][.5mm]{new HMXB?} \\  
\noalign{\smallskip}\hline\noalign{\smallskip}                       
07 & 8.1 & 073 & 114 & 35 & AX\,J0058--720, RX\,J0057.8--7202 & HMXB?, P \\
\noalign{\smallskip}\hline\noalign{\smallskip}                       
08 & 7.3 & 074 & 136 & 36 & RX\,J0057.8--7207 & HMXB?, P \\
\noalign{\smallskip}\hline\noalign{\smallskip}                       
09 & 4.3 & 076 & --  & 38 & RX\,J0058.2--7231 & HMXB Be \\
\noalign{\smallskip}\hline\noalign{\smallskip}                       
10 & 2.9 &     &     &    &     &       \\
10a& 4.7 & \raisebox{1.5mm}[.5mm][.5mm]{081} & \raisebox{1.5mm}[.5mm][.5mm]{218} & \raisebox{1.5mm}[.5mm][.5mm]{--} & \raisebox{1.5mm}[.5mm][.5mm]{RX\,J0059.3--7223} & \raisebox{1.5mm}[.5mm][.5mm]{XRB?}  \\
\noalign{\smallskip}\hline\noalign{\smallskip}                       
11 & 6.6 & 088 & 123 &--  & RX\,J0100.2--7204 & XRB? AGN?  \\
\noalign{\smallskip}\hline\noalign{\smallskip}                       
12 & --  & --  & --  &--  & XMMU\,J010030.2--722035 & new HMXB? \\
\noalign{\smallskip}\hline\noalign{\smallskip}                       
13 & 4.8 &     &     &    &    &   \\
13a& 9.8 & \raisebox{1.5mm}[.5mm][.5mm]{093} & \raisebox{1.5mm}[.5mm][.5mm]{132} & \raisebox{1.5mm}[.5mm][.5mm]{42} & \raisebox{1.5mm}[.5mm][.5mm]{RX\,J0101.0--7206} & \raisebox{1.5mm}[.5mm][.5mm]{HMXB Be} \\
\noalign{\smallskip}\hline\noalign{\smallskip}                       
14 & 6.3 & --  &     &    &    &      \\
14a& 6.3 & \raisebox{1.5mm}[.5mm][.5mm]{095} & \raisebox{1.5mm}[.5mm][.5mm]{159} & \raisebox{1.5mm}[.5mm][.5mm]{43} & \raisebox{1.5mm}[.5mm][.5mm]{RX\,J0101.3--7211} & \raisebox{1.5mm}[.5mm][.5mm]{HMXB Be, P} \\
\noalign{\smallskip}\hline\noalign{\smallskip}                       
15 & 5.3 &     &     &    &      &      \\
15a& 7.4 & \raisebox{1.5mm}[.5mm][.5mm]{096} & \raisebox{1.5mm}[.5mm][.5mm]{121} & \raisebox{1.5mm}[.5mm][.5mm]{44} & \raisebox{1.5mm}[.5mm][.5mm]{RX\,J0101.6--7204} & \raisebox{1.5mm}[.5mm][.5mm]{HMXB?}  \\
\noalign{\smallskip}\hline\noalign{\smallskip}                       
16 & 4.6 & 101 & 143 & 49 & AX\,J0103--722, SAX\,J0103.2--7209 & HMXB Be, P \\
\noalign{\smallskip}\hline\noalign{\smallskip}                       
17 & 2.6 & 105 & 106 & 50 & RX\,J0103.6--7201 & HMXB? \\
\noalign{\smallskip}\hline\noalign{\smallskip}                       
18 & 6.9 & --  & 120 & 55 & RX\,J0105.9--7203 & HMXB? \\
\noalign{\smallskip}
\hline
\end{tabular}       

\vspace{1.5mm}
Notes to column No 16:
Distance to ROSAT source.

\vspace{.5mm}
Notes to columns No 17 -- 19:
Entry numbers in ROSAT HRI catalogue \citep{2000A&AS..147...75S},
ROSAT PSPC catalogue \citep{2000A&AS..142...41H}, and  
\citet{2000A&A...359..573H}.

\vspace{.5mm}
Notes to column No 20:
YIT2000: \citet{2000ApJS..128..491Y}.

\vspace{.5mm}
Notes to column No 21:
HMXB: High mass X-ray binary, 
HMXB?: HMXB candidate, 
XRB?: X-ray binary candidate, 
Be: Be system, P: pulsar. 
\end{table*}

\begin{figure}
\centering
\includegraphics[width=6.5cm,angle=270,bb=40 75 558 787,clip]{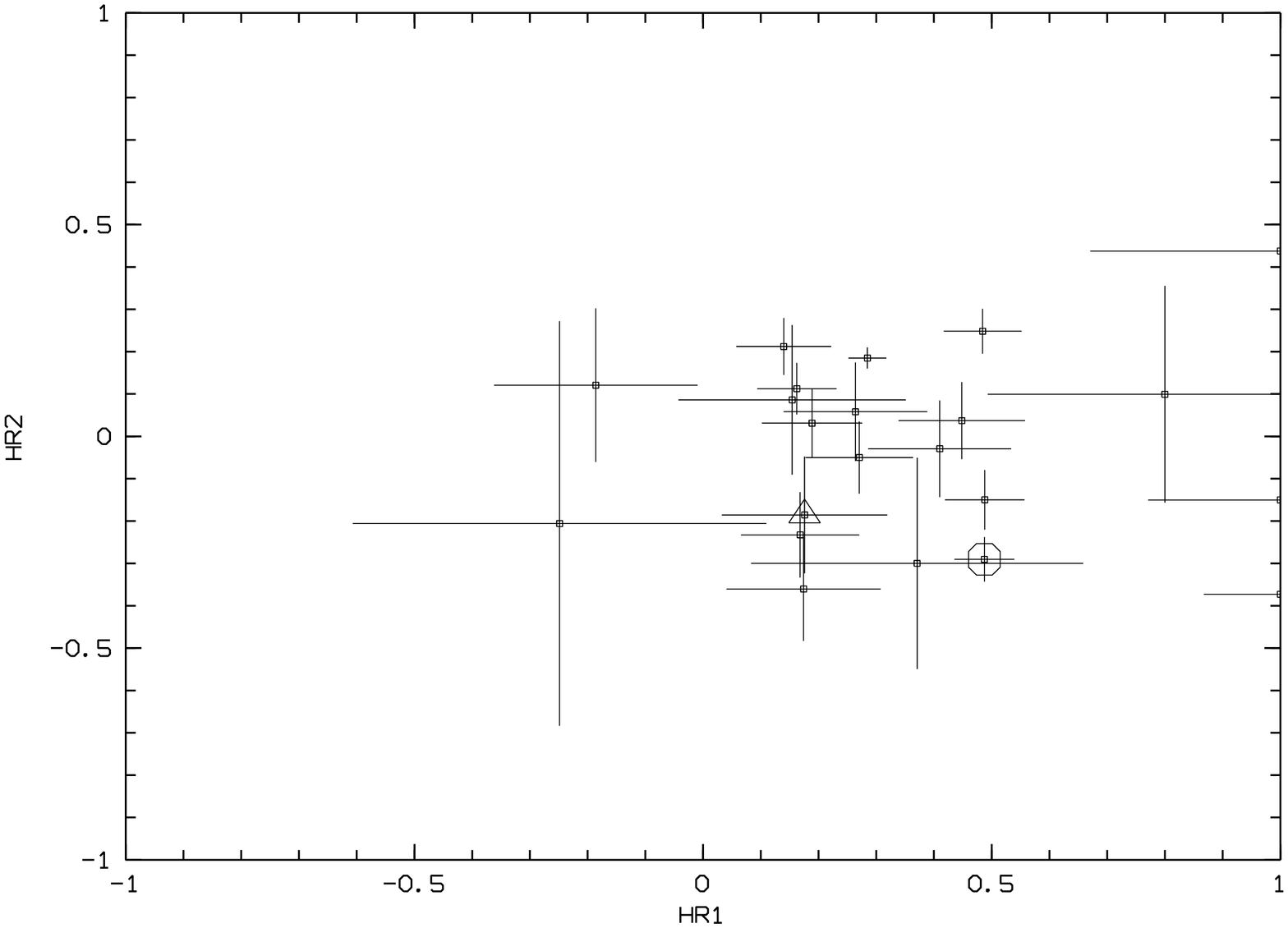}\\*[2mm]
\includegraphics[width=6.5cm,angle=270,bb=40 75 558 787,clip]{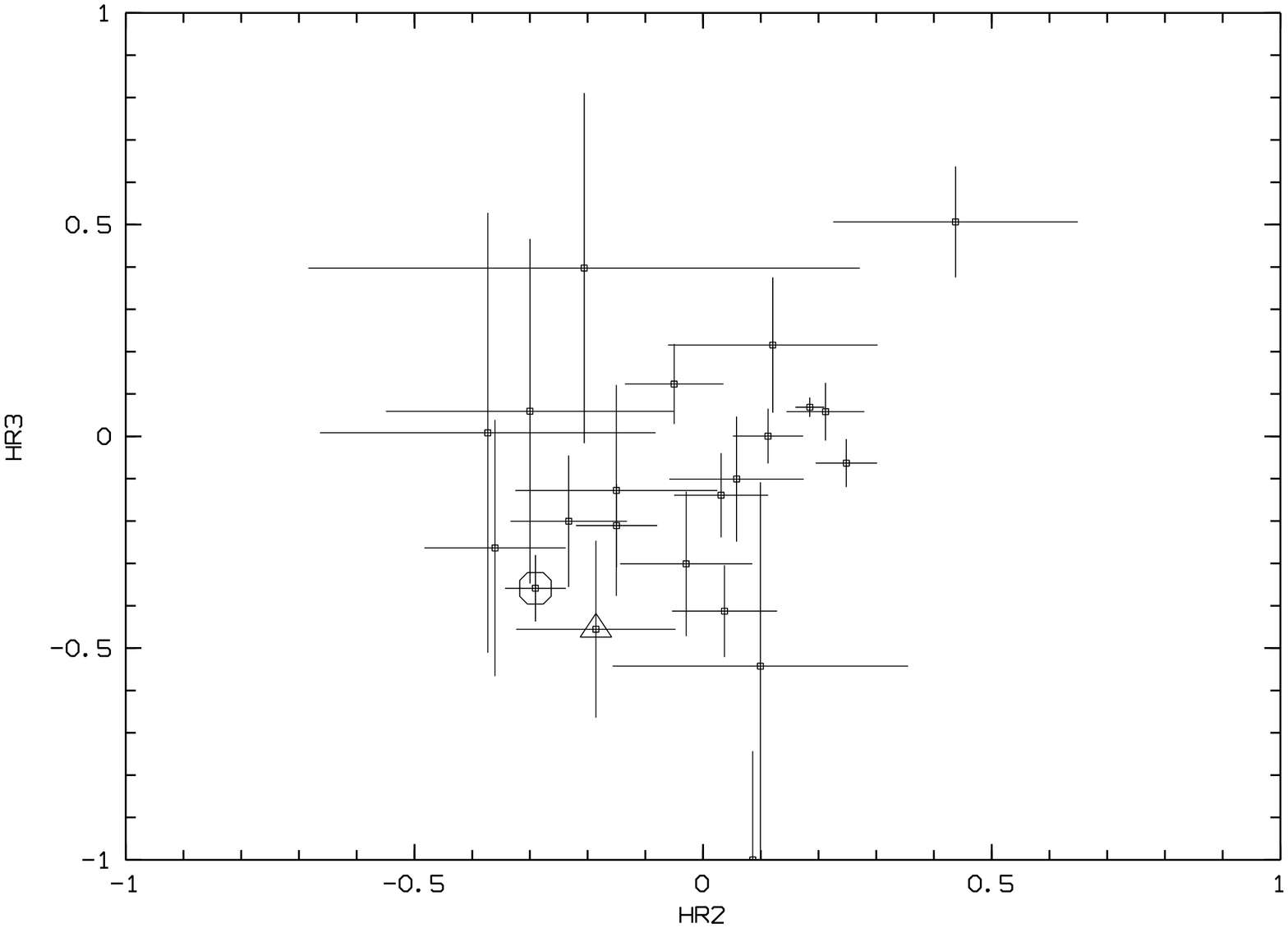}
\caption{\label{hr1hr2hr3}
HR1 plotted over HR2 and HR2 over HR3 with errors 
for all the sources in Table \ref{xbtab}
except for SMC\,X-2. Circle is used to mark the source No 5 which is an AGN, 
and triangle for source No 11 which is either an XRB or an AGN.
}
\end{figure}

All the data were processed with the XMM-Newton Science Analysis System 
(SAS) version 5.3.3.
For source detection, the events were separated into four energy bands: 
B$_{1}$ = 0.3 -- 1.0~keV, B$_{2}$ = 1.0 -- 2.0~keV, B$_{3}$ = 2.0 -- 4.5~keV, 
and B$_{4}$ = 4.5 -- 10.0~keV. In all these bands, images were created
and source detection was performed using the sliding window and maximum 
likelihood methods of the SAS. Detections with likelihood of existence 
(ML) higher than 10.0 were accepted as real sources. This corresponds to 
the probability $P = 1 - {\rm exp}(-{\rm ML}) = 0.999955$ for the existence 
of the 
source. Hardness ratios were computed using the source counts in different 
bands:
\begin{equation}\label{hr123}
{\rm HR}i = \frac{{\rm B}_{i+1} - {\rm B}_{i}}{{\rm B}_{i+1} + {\rm B}_{i}}
\end{equation}
for $i = 1, 2, 3$. 
The values of HR1, HR2, and HR3 are shown in two diagrams in 
Fig.\,\ref{hr1hr2hr3}. In most cases, X-ray binaries in the SMC or 
Active Galactic Nuclei (AGNs) behind the SMC have absorbed spectra and 
therefore show positive values for HR1.
As can be seen in the second diagram, HR2 and HR3
which compare the events in the energy bands above 1.0~keV, cluster around
zero. A large fraction of the sources (90\%) is located in a region with 
$-0.4 <$ HR2 $< 0.3$ and $-0.6 <$ HR3 $< 0.5$.

The detected sources were cross-correlated with catalogues of 
Einstein sources \citep{1992ApJS...78..391W} as well as 
ROSAT sources detected by PSPC \citep{1999A&AS..136...81K,2000A&AS..142...41H} 
and by HRI 
\citep{2000A&AS..147...75S} instruments. The positions of the X-ray sources 
were plotted on Digitized Sky Survey DSS2 (red) images of this field 
in order to find probable optical counterparts. The optical sources were also 
verified by cross-correlating the X-ray source list with the USNO-A2.0 
catalogue produced by the United States Naval Observatory 
\citep{1996AAS...188.5404M,1998AAS...19312003M}. 
In addition, we compared the source positions to the entries in the Optical 
Gravitational Lensing Experiment OGLE-II project list of variable sources in 
the Magellanic Clouds \citep{2001AcA....51..317Z}. For five out of 15 sources, 
correlations with an OGLE object were found. 
Finally, the source list was cross-correlated with the list of
emission line objects in the SMC (Meyssonnier \& Azzopardi, 
\citeyear{1993A&AS..102..451M}, [MA93]). The existence of an emission line 
star at the position of a hard X-ray source indicates that the source might be 
an X-ray binary system with a Be-star companion. 

The complete set of source lists will be presented in another paper. 
Here, we shall concentrate on the HMXBs and candidates in the four fields.
The results on the eighteen sources are summarised in Table \ref{xbtab}.
The table includes the X-ray 
source coordinates, 1\,$\sigma$ positional error, count rate, 
likelihood for
detection in the total band, flux, hardness ratios (see Eq.\,(\ref{hr123})),
pulse period with 1\,$\sigma$ error (Sect.\,\ref{timspec}), and 
identifications. 
Correlations with ROSAT sources, OGLE objects, and emission line objects 
\citep{1993A&AS..102..451M} can be found as well. The sources are sorted
by RA and Dec (J2000.0), and the entry numbers are used in the following.
The positional errors which are given in this table are statistical errors. 
The systematic error of the X-ray position 
is about 3\arcsec\ -- 4\arcsec\ \citep{2002A&A...382..522B}.
In order to calculate the flux, model parameters resulting from 
the spectral analysis (see Sect.\,\ref{timspec}) 
were used for all sources.

In the following, the source number in the ROSAT HRI catalogue of the SMC 
\citep{2000A&AS..147...75S} is given as RH\,NNN, and the number in the ROSAT 
PSPC catalogue \citep{2000A&AS..142...41H} as RP\,NNN. The entry number in 
the list of Haberl \& Sasaki (\citeyear{2000A&A...359..573H}, [HS2000]) is  
also mentioned using the format [HS2000]\,NN (also see Table \ref{xbtab}).

\subsection{Timing and spectral analysis}\label{timspec}

\begin{table*}
\caption[]{\label{spectab} Spectral parameters for sufficiently bright sources.}
\begin{tabular}{clcccllccc}
\hline\noalign{\smallskip}
\multicolumn{1}{c}{1} & \multicolumn{1}{c}{2} & \multicolumn{1}{c}{3} &
\multicolumn{1}{c}{4} & \multicolumn{1}{c}{5} & \multicolumn{1}{c}{6} &
\multicolumn{1}{c}{7} & \multicolumn{1}{c}{8} & \multicolumn{1}{c}{9} & 
\multicolumn{1}{c}{10} \\
\hline\noalign{\smallskip}
No & \multicolumn{1}{c}{ID} & $\Gamma$ & $N_{\rm H}$ & $kT$ & 
\multicolumn{1}{c}{Abund.} & 
\multicolumn{1}{c}{Thermal} & $\chi^{2}$ & dof & No \\
& & & [cm$^{-2}$] & [keV] & \multicolumn{1}{c}{(solar)} 
& \multicolumn{1}{c}{model} & & & \\
\noalign{\smallskip}\hline\noalign{\smallskip}
05 & 2E\,0055.6--7241 & $2.59^{+0.16}_{-0.33}$ & $5.3^{+1.1}_{-0.8} 
\times 10^{21}$ & -- & \multicolumn{1}{c}{--} & -- & 41.4 & 51 & 2 \\
\noalign{\smallskip}\hline\noalign{\smallskip}
06 & XMMU\,J005735.7--721932 & $1.42^{+0.25}_{-0.20}$ & $3.6^{+1.9}_{-1.5} 
\times 10^{21}$ & -- & \multicolumn{1}{c}{--} & -- & 84.6 & 63 & 4 \\
\noalign{\smallskip}\hline\noalign{\smallskip}
07 & AX\,J0058--720          & $1.01^{+0.11}_{-0.11}$ & $3.4^{+5.9}_{-3.4} 
\times 10^{20}$ & -- & \multicolumn{1}{c}{--} & -- & 91.6 & 65 & 3 \\
\noalign{\smallskip}\hline\noalign{\smallskip}
08 & RX\,J0057.8--7207       & $0.97^{+0.08}_{-0.07}$ & $3.0^{+0.8}_{-0.4} 
\times 10^{21}$ & -- & \multicolumn{1}{c}{--} & -- & 161.9 & 133 & 3 \\
\noalign{\smallskip}\hline\noalign{\smallskip}
10 & RX\,J0059.3--7223       & $1.46^{+0.12}_{-0.13}$ & $1.8^{+0.7}_{-0.6} 
\times 10^{21}$ & -- & \multicolumn{1}{c}{--} & -- & 117.8 & 99 & 5 \\
\noalign{\smallskip}\hline\noalign{\smallskip}
11 & RX\,J0100.2--7204       & $2.00^{+0.41}_{-0.26}$ & $1.7^{+0.9}_{-1.0} 
\times 10^{21}$ & -- & \multicolumn{1}{c}{--} & -- & 28.9 & 26 & 3 \\
\noalign{\smallskip}\hline\noalign{\smallskip}
14 & RX\,J0101.3--7211       & $1.14^{+0.18}_{-0.13}$ & $3.3^{+2.4}_{-1.1} 
\times 10^{21}$ & $0.20^{+0.09}_{-0.06}$ & 
\multicolumn{1}{c}{$0.11^{+0.10}_{-0.11}$} & MEKAL 
& 82.6 & 56 & 2 \\
\noalign{\smallskip}\hline\noalign{\smallskip}
15 & RX\,J0101.6--7204       & $1.73^{+0.15}_{-0.17}$ & $8.6^{+4.9}_{-5.6} 
\times 10^{20}$ & -- & \multicolumn{1}{c}{--} & -- & 68.0 & 58 & 6 \\
\noalign{\smallskip}\hline\noalign{\smallskip}
16 & AX\,J0103--722          & $1.08_{-0.19}^{+0.12}$ & $1.9_{-1.7}^{+1.9} 
\times 10^{21}$ & $0.27_{-0.07}^{+0.08}$ & 
\multicolumn{1}{c}{$0.31_{-0.17}^{+0.43}$} & MEKAL 
& 172.7 & 155 & 3\\
\noalign{\smallskip}\hline\noalign{\smallskip}
17 & RX\,J0103.6--7201 (Model 1) & $0.72_{-0.07}^{+0.06}$ & $1.7_{-0.9}^{+1.0} 
\times 10^{21}$ & $0.27_{-0.04}^{+0.03}$ & 
\multicolumn{1}{c}{$0.77_{-0.28}^{+0.17}$} & MEKAL 
& 295.1 & 240 & 3 \\
\noalign{\smallskip}
& \multicolumn{1}{r}{(Model 2)} & $0.71_{-0.06}^{+0.05}$ & $2.3_{-0.8}^{+0.8} \times 10^{21}$ & $0.32_{-0.05}^{+0.05}$ & $1.07_{-0.18}^{+0.57}$ (O) & VMEKAL & 272.9 & 238 & 3 \\
 & & & & & $1.61_{-0.66}^{+1.40}$ (Ne) & & & & \\
\noalign{\smallskip}\hline
\end{tabular}

\vspace{1.5mm}
Notes to column No 4:
Additional $N_{\rm H}$ to Galactic foreground $N_{\rm H, Gal}$ 
(see Eqs.\,(\ref{power}) and (\ref{powme})).

\vspace{.5mm}
Notes to column No 9:
Degrees of freedom.

\vspace{.5mm}
Notes to column No 10:
Number of used spectra.
\end{table*}

As the very first step of data analysis, we checked the EPIC PN data for 
incorrect
time information. It has been reported that in some cases, there are time jumps 
of 1~s in the EPIC PN data which were not corrected in the SAS processing.  
Since we didn't find any event which indicated such a time jump, we could 
proceed without any countermeasure.

After selecting the events for each source, they were
analysed using the XANADU software package distributed by the High Energy 
Astrophysics Science Archive Research Center (HEASARC). It contains the
packages XRONOS for timing analysis and XSPEC for spectral fitting.

Based on EPIC PN data, period search was carried out with XRONOS after 
correcting the photon arrival times for solar system barycentre. 
If a peak    
was found in the power spectrum indicating pulsations, a more detailed epoch  
folding search was performed around the preliminary value. 
Once we got the rough value for the pulse period, the $\chi^{2}$ 
distribution around this value was fitted with a Lorentz profile and the 
maximum of the Lorentz profile 
was determined together with the 1~$\sigma$ error. Finally, folded light 
curves were created in three energy bands: B$_{1}$ = 0.3 -- 1.0~keV, 
B$_{2}$ = 1.0 -- 2.0~keV, B$_{3+4}$ = 2.0 -- 10.0~keV. 
In addition, the ratio of the count rates in the harder band to the softer 
band (B$_{2}$/B$_{1}$ and B$_{3+4}$/B$_{1+2}$, with B$_{1+2}$ = 0.3 -- 2.0~keV)
were computed to illustrate the changes in the hardness ratios with pulse
phase. Note that these hardness ratios are different to the numbers defined
in Eq.\,(\ref{hr123}).

Except for sources which were too faint, spectra were extracted for each 
source. These spectra were modelled with a power law component together with 
the fixed Galactic foreground absorbing column density of 
$N_{\rm H, Gal} = 5.74 \times 10^{20}$~cm$^{-2}$ \citep{1990ARA&A..28..215D}
and a free column density $N_{\rm H}$: 
\begin{equation}\label{power}
S_{1}(E) = e^{-\sigma(E)\,N_{\rm H\,Gal}} \times e^{-\sigma(E)\,N_{\rm H}} \times K \times E^{-\Gamma},
\end{equation}
$E$ being the energy in [keV], $\Gamma$ the photon index, and $K$ the 
normalisation.
In some cases there was a deviation of the observed spectrum 
from a power law spectrum, suggesting the existence of an additional soft 
component. 
Since the spectra show features indicating emission 
lines, the soft component was modelled as thermal plasma emission.
This thermal emission presumably arises from circumstellar matter, and the 
absorption must be negligibly low in comparison to the absorption of
the hard X-ray emission from the neutron star. Moreover, if the column density 
was high, the soft emission would be absorbed and thus not detectable. 
Using the MEKAL model in XSPEC 
\citep{1985A&AS...62..197M,1986A&AS...65..511M,1992SRON..........K,1995ApJ...438L.115L} 
for the thermal component without additional absorbing 
column density, the spectrum can be written as
\begin{eqnarray}\label{powme}
S_{2}(E) & = & e^{-\sigma(E)\,N_{\rm H\,Gal}} \times (e^{-\sigma(E)\,N_{\rm H}} \times K \times E^{-\Gamma} \nonumber \\
& & + S_{\rm MEKAL}(T,{\rm Abund.})).
\end{eqnarray}
$S_{\rm MEKAL}(T)$ is the MEKAL model spectrum with a temperature corresponding
to $kT$ in [keV] and elemental abundances with respect to solar. 
We also performed a fit with a blackbody component instead of the MEKAL
model. Although for fainter sources, no significant difference was found 
in the fits, for bright sources like No 16 (Fig.\,\ref{J0103-7209})
and No 17 (Fig.\,\ref{J0103-7201}), the blackbody fit results in higher 
$\chi^{2}$: 231.5 for 156 degrees of freedom for source No 16 and
322.5 for 241 degrees of freedom for source No 17 (compare
to Table \ref{spectab}).
This is because the blackbody model fits the low energy tail of the spectrum,
but does not account for the peaks around 0.6~keV and 0.9~keV, which might 
indicate emission lines from highly ionized oxygen and neon, as well as iron
lines.
For sources which were bright enough to obtain a significant spectral
fit, the results are shown in figures and the model parameters yielding the 
best fit results are listed in the Table \ref{spectab} together with 
1\,$\sigma$ errors.
Moreover, for source No 17 which is very bright and shows emission lines
most prominently, we also used the VMEKAL  
instead of the MEKAL model. In this more elaborate model, the 
abundance for each of the element is a variable fit parameter.
We obtained an improved fit with high oxygen and neon abundances, whereas
the other elements have values below solar, not differing significantly from 
zero.

From the spectral models for the emission we were able to estimate the flux 
of the sources in the ROSAT band (0.1 -- 2.4~keV). 
The flux was calculated from the fitted models, except for 
four sources which were too faint:
For No 02, a power law spectrum with $\Gamma = 0.7$ and 
$N_{\rm H} = 0.0 \times 10^{21}$~cm$^{-2}$
\citep{2001PASJ...53..227Y} was assumed to estimate the flux upper limit. 
For No 01, 09, and 12, $\Gamma = 1.0$ and   
$N_{\rm H} = 1.0 \times 10^{21}$~cm$^{-2}$ were adopted. 
The resulting luminosity 
was used to create a long term light curve of all the ROSAT and the new 
XMM-Newton data. In the light curves, crosses are  
used for ROSAT PSPC data, triangles for ROSAT HRI data, and dots for XMM-Newton
EPIC data. Upper limits determined from ROSAT observations are plotted as 
arrows. For the distance to the SMC, a mean value of 60~kpc was 
assumed \citep[see review by][]{1999IAUS..190..569V}. 

\section{Comments on individual HMXBs and candidates}

In this section, we present the results on individual sources.
All sources, which were detected in the four data sets and were proven to be 
HMXBs or candidates, are listed in Table \ref{xbtab}. 

\subsection{Source No 1: RX\,J0051.7--7341}

\object{RX\,J0051.7--7341} which has been suggested as an XRB candidate by
\citet{1999A&AS..136...81K} was only detected in MOS1/2 data. In the PN data
the source was located on a bad column. It is faint,
so neither spectral nor timing analysis was performed for this source. 
The PSPC count rate of $1.64 \times 10^{-3} \pm 0.68 \times 10^{-3}$~s$^{-1}$ 
\citep{1999A&AS..136...81K} during the ROSAT observation corresponds to
XMM-Newton MOS (medium filter) count rate of about
$8 \times 10^{-3}$~s$^{-1}$. This means that the 
luminosities of the source during the ROSAT and XMM-Newton observations 
were comparable (see Table \ref{xbtab}).

\subsection{Source No 2: SMC\,X-2}

\object{SMC\,X-2} was one of the first three X-ray sources which were 
discovered in the SMC \citep{1978ApJ...221L..37C}. It was also detected in the 
HEAO\,1 A-2 experiment \citep{1979ApJS...40..657M}, but not in the Einstein IPC
survey \citep{1981ApJ...243..736S}. In ROSAT observations, this transient 
source was detected only once \citep{1996A&A...312..919K}. It is thought to be 
a Be/XRB, since a Be-star was found as its optical counterpart 
\citep{1979MNRAS.186P..43M}. In early 2000, the RXTE All-Sky Monitor detected 
an outburst at the position of SMC\,X-2 \citep{2001ApJ...548L..41C} 
and a pulse period of 2.374$\pm$0.007~s was determined
\citep{2000IAUC.7402....3C,2000IAUC.7441....2T}.

In the XMM-Newton data (Obs.\ ID 00842008), there was no detection with 
ML $>$ 10 (see Sect.\,\ref{soudet}) 
at the position of SMC\,X-2 which was apparently in low luminosity 
state during the XMM-Newton observation. Therefore, we performed source 
detection using the maximum likelihood routine at the position of the 
optical counterpart (SIMBAD):
RA = 00$^h$ 54$^m$ 33.4$^s$, Dec = --73\degr\ 41\arcmin\ 04\arcsec\ (J2000.0).
Since we set the ML limit lower, the source was detected with a 
likelihood of ML = 3.4.
The 3~$\sigma$ upper limit count rate obtained from the ML source detection  
routine is $2.33 \times 10^{-3}$~s$^{-1}$. 
The source counts were highest in the B$_{3}$ 
band (2.0 -- 4.5~keV). 
In order to estimate the flux upper limit, spectral parameters derived by 
\citet{2001PASJ...53..227Y} from the ASCA spectrum during the outburst were 
used: 
Photon index $\Gamma = 0.7$ for a power law spectrum absorbed by a column 
density of $N_{\rm H} < 1.0 \times 10^{21}$~cm$^{-2}$. This results in an 
upper limit for the
un-absorbed flux of $1.5 \times 10^{-14}$~erg~cm$^{-2}$~s$^{-1}$, corresponding
to $L_{\rm X} = 6.5 \times 10^{33}$~erg~s$^{-1}$
(0.3 -- 10.0~keV) during the XMM-Newton observation in Oct.\ 2001.

\subsection{Source No 3: RX\,J0054.9--7226}

\begin{figure}
\centering
\includegraphics[width=5.3cm,angle=270]{h4167f3a.ps}\\*[2mm]
\includegraphics[width=5.3cm,angle=270]{h4167f3b.ps}\\*[2mm]
\hspace{-4.7mm}\includegraphics[width=5.65cm,angle=270,bb=40 50 560 772,clip]{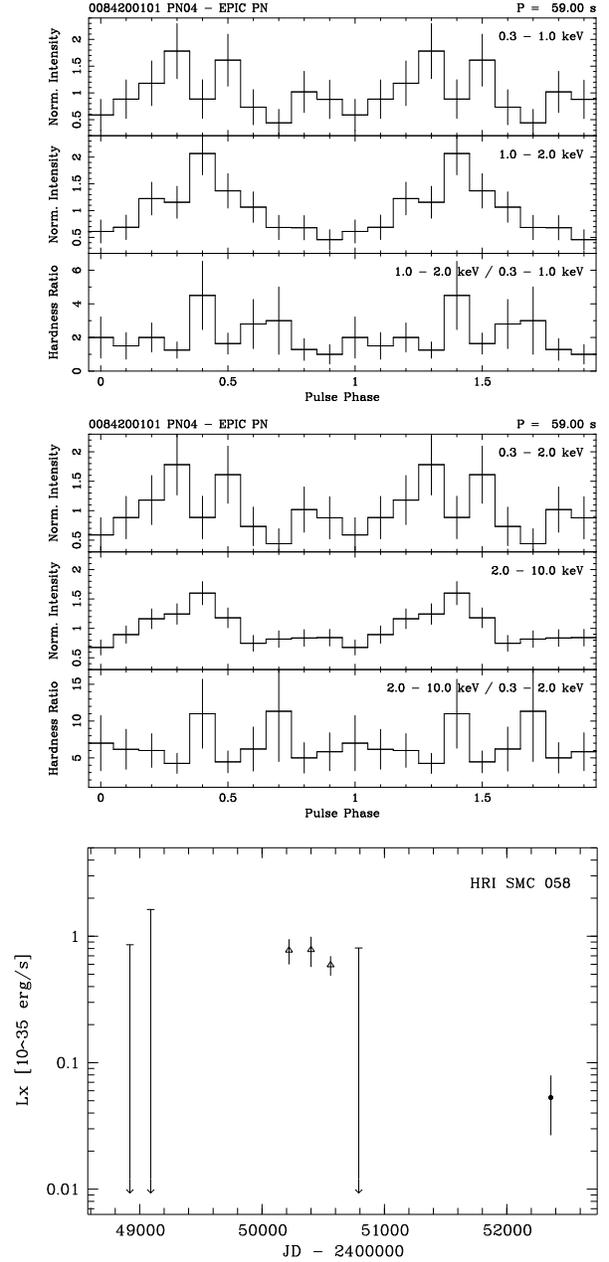}
\caption{\label{J0054-7226}
Folded light curves and long term light curve of RX\,J0054.9--7226 
(source No 3).
The hardness ratio is the ratio between the count rates in harder band and 
the count rates in softer band (Sect.\,\ref{timspec}). See text for the symbols
used for the long term light curve.
}
\end{figure}                                  

\object{RX\,J0054.9--7226} is known to be an X-ray binary pulsar with a pulse 
period of 58.969$\pm$0.001~s 
\citep{1998IAUC.6818....1M,1998A&A...338L..59S} and is the only source in our
sample, for which the orbital period has been measured: 65~d 
\citep{1999AAS...194.5218L}.
In the timing analysis of the new XMM-Newton data, the pulse period was 
verified to be 59.00$\pm$0.02~s. The folded light curves show variations 
especially above 1.0~keV, and there is no significant change in hardness 
ratios (Fig.\,\ref{J0054-7226}). As can be seen in the long term light curve, 
compared to ROSAT data, the source was observed in low luminosity state. 
Due to the low flux, the statistics of the spectrum were not high enough and 
the spectrum is thus 
not discussed here. However, the results of the
spectral analysis was used to estimate the flux of the source 
(see Table \ref{xbtab}). The 
optical counterpart, a Be-star, is identified with the variable star 
OGLE\,00545617--7226476 \citep{2001AcA....51..317Z}.

\subsection{Source No 4: XMMU\,J005605.2--722200 = 2E\,0054.4--7237?}

\begin{figure}
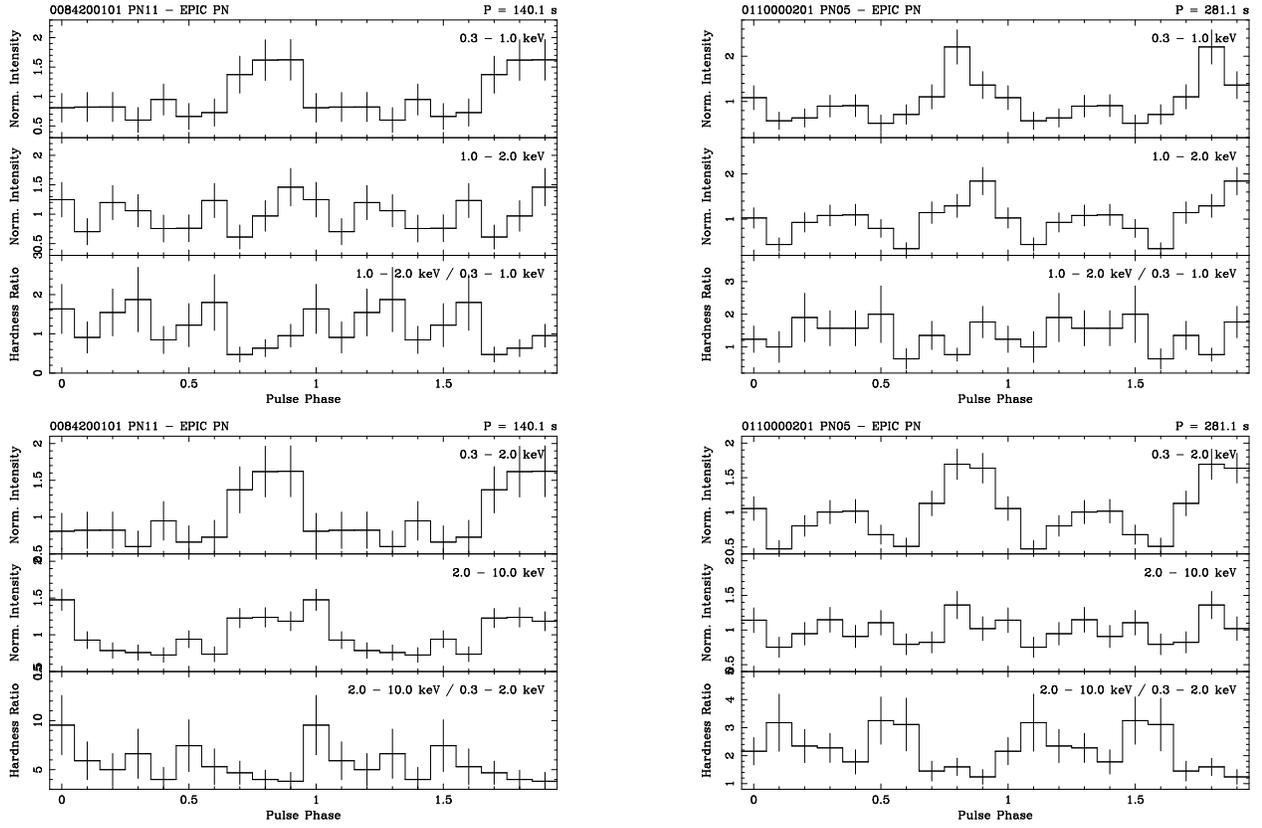

\centering
\includegraphics[width=5.3cm,angle=270]{h4167f4a.ps}\\*[2mm]
\includegraphics[width=5.3cm,angle=270]{h4167f4b.ps}\\*[2mm]
\caption{\label{J0056-7222}
Folded light curves of XMMU\,J005605.2--722200 (source No 4).
Hardness ratio as in Fig\,\ref{J0054-7226}.
}
\end{figure}                                  

The error circle of the Einstein source \object{2E\,0054.4--7237} 
includes an emission line object. 
Therefore, it was suggested as a Be/XRB candidate ([HS2000]).
In the XMM-Newton data, a source consistent with the position of the emission
line object was detected (\object{XMMU\,J005605.2--722200}) and pulsations
from this source was discovered. XMMU\,J005605.2--722200 is most
likely consistent with \object{2E\,0054.4--7237}.   
The period is 140.1$\pm$0.3~s. As can be seen in Fig.\,\ref{J0056-7222}, 
the pulses in the soft band are narrower than in the harder band.

\subsection{Source No 5: 2E\,0055.6--7241}\label{souno5}

\begin{figure}
\centering
\hspace{.7mm}\includegraphics[width=5.2cm,angle=270,clip]{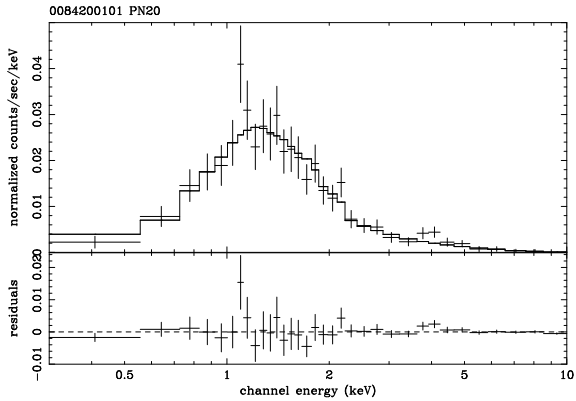}
\caption{\label{J0057-7225}
Spectrum of 2E\,0055.6--7241 (source No 5).
}
\end{figure}                                  

\object{2E\,0055.6--7241} had been suggested as an XRB candidate by
\citet{1999A&AS..136...81K}. 
Timing analysis revealed no pulsations of the X-ray source.
Also on longer timescales no flux change was verified:
The ROSAT PSPC count rate was 
$7.48 \times 10^{-3} \pm 0.62 \times 10^{-3}$~s$^{-1}$ 
\citep{1999A&AS..136...81K}, corresponding to a count rate of 
$8 \times 10^{-2}$~s$^{-1}$ for XMM-Newton 
EPIC PN (thin1 filter). This value is
similar to the count rate of the XMM-Newton observation, 
which is $8.79 \times 10^{-2} \pm 0.42 \times 10^{-2}$~s$^{-1}$.
The X-ray spectrum is shown in Fig.\,\ref{J0057-7225}. It has a photon  
index of $\Gamma = 2.59^{+0.16}_{-0.33}$, which is higher than for 
other sources of our sample, and highest absorbing column 
density of $N_{\rm H} = 5.3^{+1.1}_{-0.8} \times 10^{21}$~cm$^{-2}$
(also see Table \ref{spectab}). 
A difference to other sources is 
also seen in the hardness ratios, as the 
source has a relatively high HR1 and lower values
of HR2 and HR3 (HR1 = +0.49$\pm$0.05, HR2 = --0.29$\pm$0.05, 
HR3 = --0.36$\pm$0.08, also see Fig.\,\ref{hr1hr2hr3}).
The high absorption makes HR1 positive, 
whereas HR2 and HR3 are negative
due to steeper power law spectrum.

On the DSS2 (red) image, there is a source 
at the X-ray position, which coincides with the variable object
OGLE\,00571981--72253375 \citep{2001AcA....51..317Z} 
with $B = 19.7$ and $R = 17.8$ (USNO-A2.0\,0150-00625436), i.e.\
$B - R = 1.9$.
\citet{2000A&AS..147...75S} have shown, that all the HMXBs and candidates
in the SMC HRI catalogue have $14 < R < 18$ and $-2 < B - R < 3$,
whereas e.g.\ AGNs have $R > 16$ and $B - R > 0$.
Both the optical magnitudes and the
X-ray spectra indicate that this source
might as well be an AGN. 
Spectroscopy of the optical counterpart by \citet{2003aph0301617....D} 
showed that this object is a $z$ = 0.15 quasar located
behind the SMC.

\subsection{Source No 7: AX\,J0058--720}  

\begin{figure}
\centering
\includegraphics[width=5.3cm,angle=270]{h4167f6a.ps}\\*[2mm]
\includegraphics[width=5.3cm,angle=270]{h4167f6b.ps}\\*[2mm]
\hspace{.7mm}\includegraphics[width=5.1cm,angle=270,clip]{h4167f6c.ps}\\*[2mm]
\hspace{-4.7mm}\includegraphics[width=5.65cm,angle=270,bb=40 50 560 772,clip]{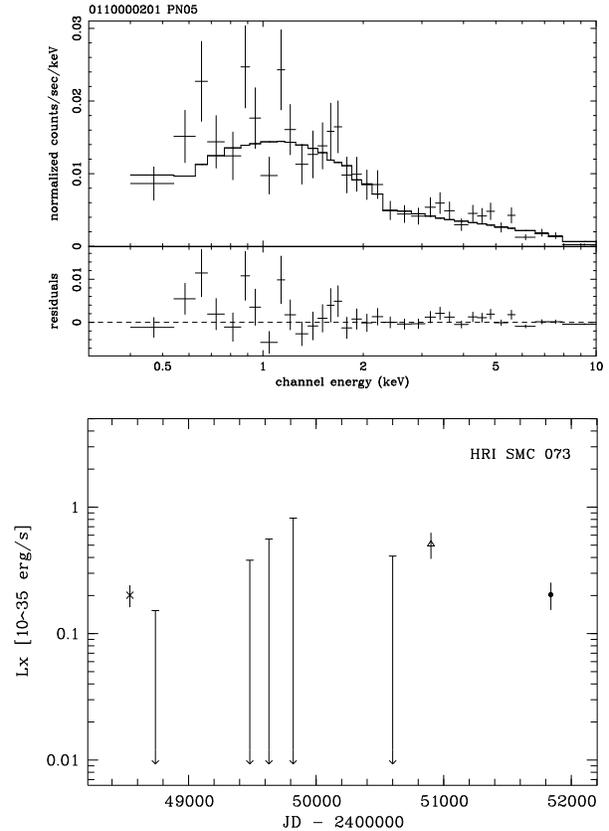}
\caption{\label{J0058-7202}
Folded light curves, spectrum, and long term light curve (0.1 -- 2.4~keV)
of AX\,J0058--720 (source No 7).  
Hardness ratio and symbols as in Fig\,\ref{J0054-7226}.
}
\end{figure}                                  

The pulse period of 
\object{AX\,J0058--720} was determined from the ASCA data as 280.4$\pm$0.3~s 
\citep{1998IAUC.6853....2Y}, which we confirmed in the XMM-Newton data: 
281.1$\pm$0.2~s. It shows strong pulses in the softer bands and its 
spectrum becomes harder during the 'off' time (Fig.\,\ref{J0058-7202}). 
The residuals of the power law fit (Table \ref{spectab} and 
Fig.\,\ref{J0058-7202}) indicate the existence of an additional soft 
component. The source has been suggested as a HMXB candidate due to the
likely optical counterpart, which is an emission line object ([HS2000]).

\subsection{Source No 8: RX\,J0057.8--7207}

\begin{figure}
\centering
\includegraphics[width=5.3cm,angle=270]{h4167f7a.ps}\\*[2mm]
\includegraphics[width=5.3cm,angle=270]{h4167f7b.ps}\\*[2mm]
\hspace{.7mm}\includegraphics[width=5.1cm,angle=270,clip]{h4167f7c.ps}\\*[2mm]
\hspace{-4.7mm}\includegraphics[width=5.65cm,angle=270,bb=40 50 560 772,clip]{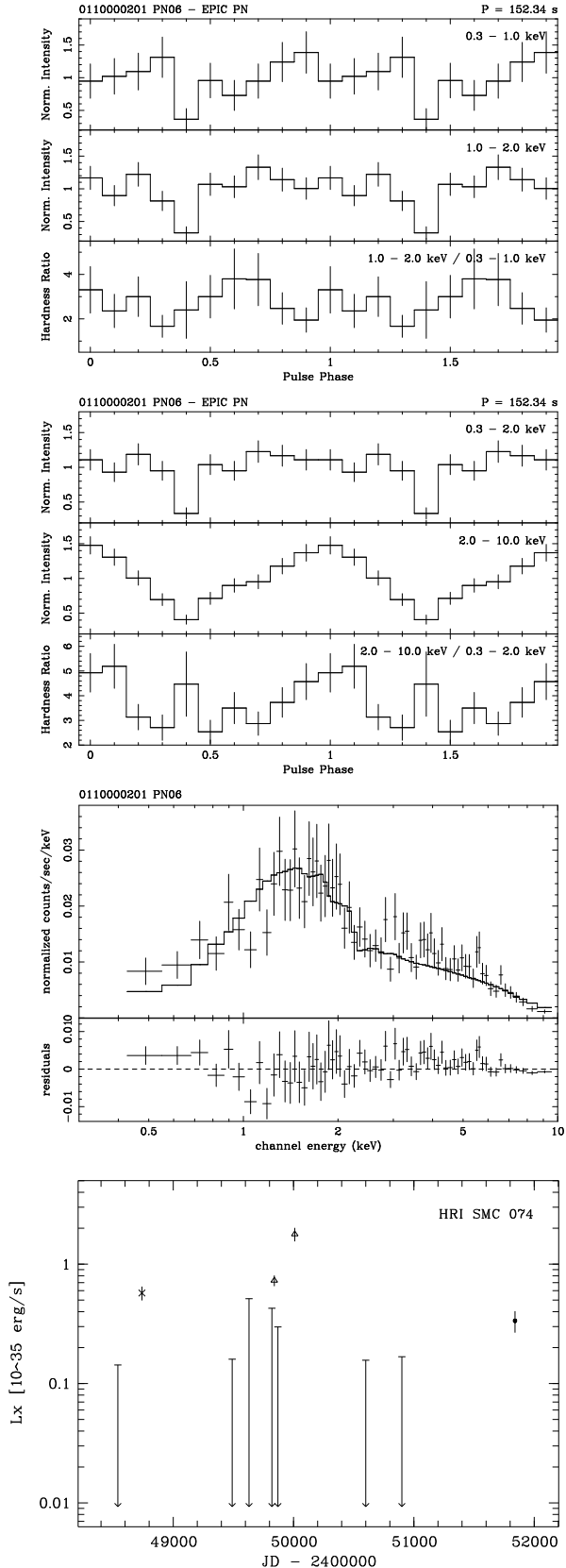}
\caption{\label{J0057-7207}  
Folded light curves, spectrum, and long term light curve (0.1 -- 2.4~keV) of 
RX\,J0057.8--7207 (source No 8).
Hardness ratio and symbols as in Fig\,\ref{J0054-7226}.
}
\end{figure}                                  

\object{RX\,J0057.8--7207} is a HMXB candidate with an emission line object 
suggested as a likely optical counterpart ([HS2000]). We discovered pulsations 
in the new XMM-Newton data and derived
a pulse period of 152.34$\pm$0.05~s. 
For this source, a period of 152.098$\pm$0.016~s was independently 
found in Chandra data by \citet{2003ApJ...584L..79M}.
The folded light curves in Fig.\,\ref{J0057-7207}
show, that there are correlated flux variations in all bands with a
significant minimum at phase 0.4. Especially in the hard band, there is a 
slow increase and fast decay. Therefore, the hardness ratio falls off at 
phase 0.2 and increases slowly after phase 0.7. 
The source spectrum is 
well reproduced by a power law spectrum (see Table \ref{spectab}) with a 
significant absorption within the SMC or the source itself.
As can be seen in the long term light curve, there was a weak flare observed by
ROSAT, whereas the XMM-Newton observation was performed in a low 
luminosity state, 5.3 times 
lower than the maximum observed by ROSAT.

\subsection{Source No 9: RX\,J0058.2--7231}

The source corresponding to the optically identified 
HMXB \object{RX\,J0058.2--7231} is very faint, so that 
no timing analysis could be performed. However, the hardness 
ratios HR1, HR2, and HR3 indicate, that this source has a hard spectrum. 
Its optical counterpart is a variable Be star in the SMC, 
OGLE\,00581258--7230485 \citep{2001AcA....51..317Z}.
From the ROSAT HRI count rate of  
$4.28 \times 10^{-3} \pm 0.48 \times 10^{-3}$~s$^{-1}$ 
\citep{2000A&AS..147...75S} we estimated the corresponding XMM-Newton 
EPIC PN (thin1 filter) count rate: $\sim2 \times 10^{-1}$~s$^{-1}$.
The source was about 3.6 times brighter when it was detected by ROSAT than
when it was observed by XMM-Newton.

\subsection{Source No 10: RX\,J0059.3--7223}

\begin{figure}
\centering
\hspace{.7mm}\includegraphics[width=5.1cm,angle=270,clip]{h4167f8a.ps}\\*[2mm]
\hspace{-4.7mm}\includegraphics[width=5.65cm,angle=270,bb=40 50 560 772,clip]{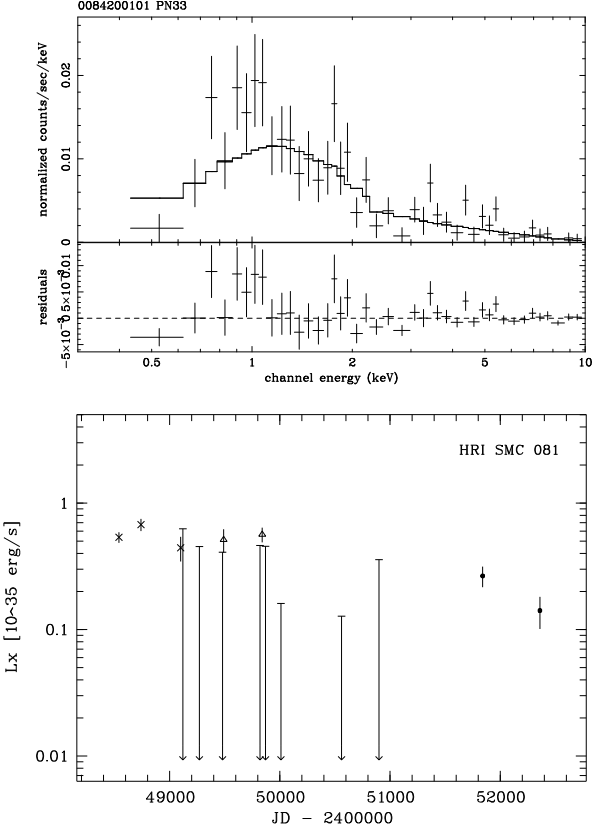}
\caption{\label{J0059-7223}
Spectrum and long term light curve (0.1 -- 2.4~keV) of RX\,J0059.3--7223 
(source No 10).
Symbols for the long term light curve as in Fig.\,\ref{J0054-7226}.}
\end{figure}                                  

\object{RX\,J0059.3--7223} has been suggested as an XRB candidate by
\citet{1999A&AS..136...81K}. 
It was observed by XMM-Newton in two pointings. 
Its spectrum mainly consists of a power law component 
typical for a HMXB with additional features 
(Fig.\,\ref{J0059-7223}). 
For this source no pulsations were detected.
At its position, there is the variable star OGLE\,00592103--7223171 
\citep{2001AcA....51..317Z}, which is 
suggested as the optical counterpart. Its magnitudes
are $B = 17.4$ and $R = 14.6$ (USNO-A2.0\,0150-00660299),
which gives $B - R = 2.8$. 
The $R$ magnitude in particular is characteristic
for a HMXB (see Sect.\,\ref{souno5}).

\subsection{Source No 11: RX\,J0100.2--7204}

\begin{figure}
\centering
\hspace{.7mm}\includegraphics[width=5.1cm,angle=270,clip]{h4167f9a.ps}\\*[2mm]
\hspace{-4.7mm}\includegraphics[width=5.65cm,angle=270,bb=40 50 560 772,clip]{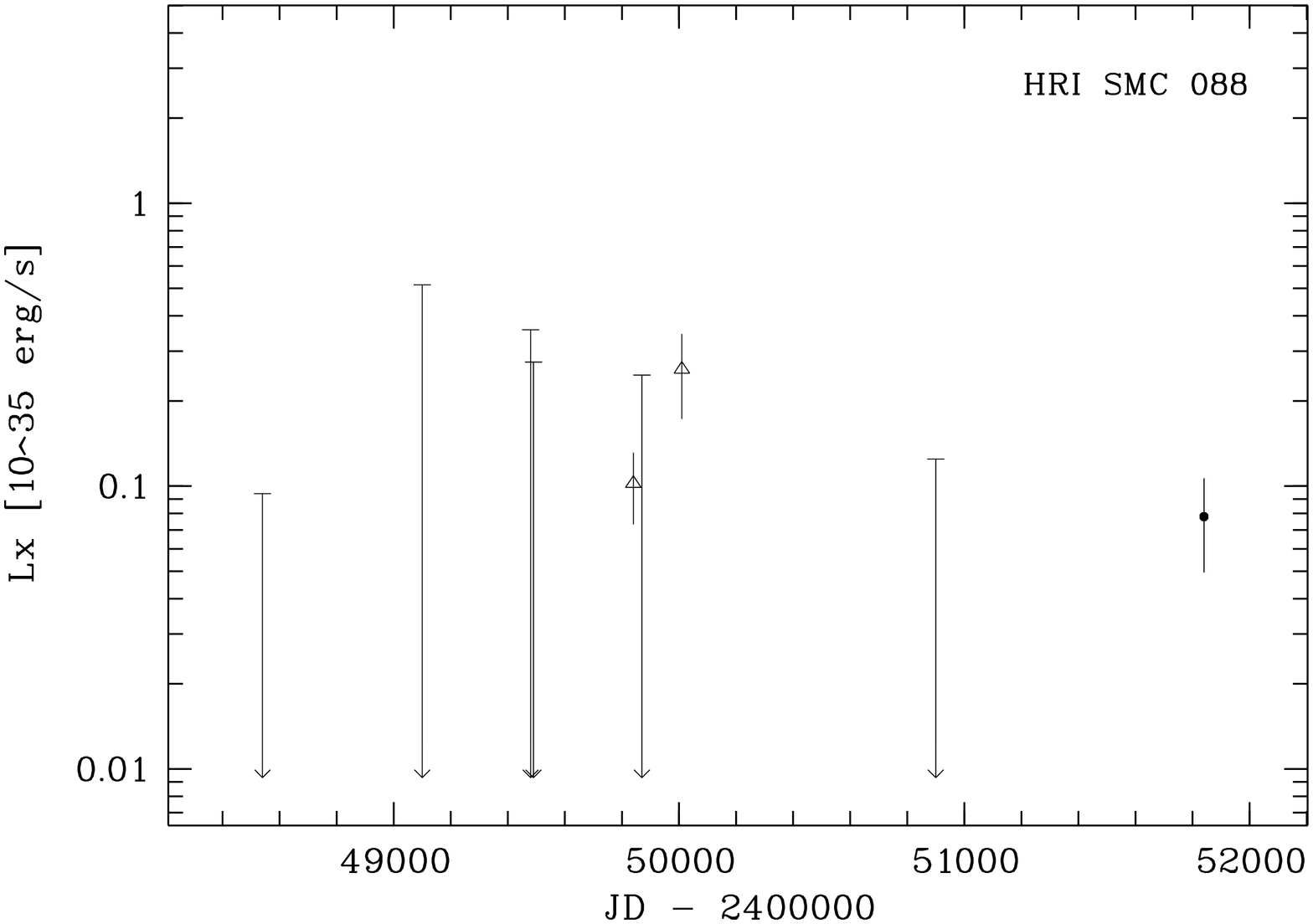}
\caption{\label{J0100-7204}    
Spectrum and long term light curve (0.1 -- 2.4~keV) of RX\,J0100.2--7204 
(source No 11).
Symbols for the long term light curve as in Fig.\,\ref{J0054-7226}.
}
\end{figure}                                  

At the position of the XMM-Newton detection corresponding to 
\object{RX\,J0100.2--7204}, a very faint object can 
be found on the DSS2 (red) image. 
However, there is no entry in the USNO-A2.0 catalogue for this source. 
We also looked for information in different catalogues using BROWSE of 
the HEASARC archive, but could not 
find the magnitudes of this optical source.
The X-ray source was suggested as an XRB candidate by 
\citet{1999A&AS..136...81K}. The spectrum of the source is a power law 
with $\Gamma = 2.00^{+0.41}_{-0.26}$ and absorbing column 
density of $N_{\rm H} = 1.7^{+0.9}_{-1.0} \times 10^{21}$~cm$^{-2}$ 
(Fig.\,\ref{J0100-7204}).
Since the probable optical counterpart is very faint and the power law
photon index is higher than for most of the other sources presented here, 
it can not be 
ruled out that this source is an AGN (also see Table \ref{spectab}). 

\subsection{Source No 13: RX\,J0101.0--7206}

\begin{figure}
\centering
\hspace{-4.7mm}\includegraphics[width=5.65cm,angle=270,bb=40 50 560 772,clip]{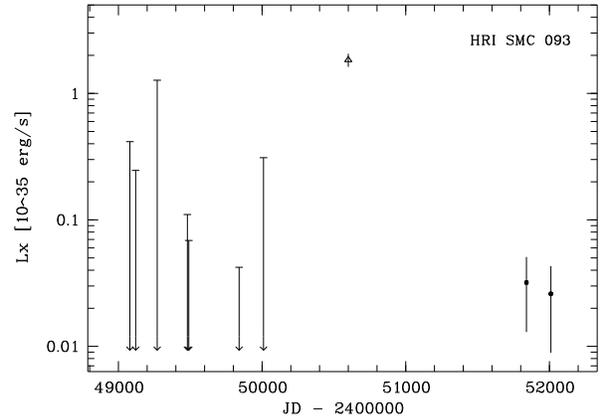}
\caption{\label{J0101-7206}    
Long term light curve (0.1 -- 2.4~keV) of RX\,J0101.0--7206 (source No 13).
Symbols as in Fig.\,\ref{J0054-7226}.
}
\end{figure}                                  

The Be/X-ray binary \object{RX\,J0101.0--7206} showed a luminosity 
of $\sim3 \times 10^{33}$~erg~s$^{-1}$ in the ROSAT band (0.1 -- 2.4\,keV) 
during two XMM-Newton observations. It was about 60 times fainter than at the 
maximum observed by ROSAT (Fig.\,\ref{J0101-7206}). 
Pulsations with a period of 304.49$\pm$0.13~s were discovered in Chandra data
\citep{2003ApJ...584L..79M}. This period could not be verified in the 
XMM-Newton observation, because the source was too faint.
\citet{2003MNRAS.338..428E} presented results on the optical analysis of 
likely counterparts, discussing two objects (No 1 and 4) in the ROSAT PSPC 
error circle. They conclude that the optical counterpart is object No 1
which is confirmed to be a Be star. This object is also the only optical source,
which can be found on the DSS image within the XMM-Newton 1~$\sigma$ error 
circle.

\subsection{Source No 14: RX\,J0101.3--7211}

\begin{figure}
\centering
\includegraphics[width=5.3cm,angle=270]{h4167f11a.ps}\\*[2mm]
\includegraphics[width=5.3cm,angle=270]{h4167f11b.ps}\\*[2mm]
\hspace{.7mm}\includegraphics[width=5.1cm,angle=270,clip]{h4167f11c.ps}\\*[2mm]   
\hspace{-4.7mm}\includegraphics[width=5.65cm,angle=270,bb=40 50 560 772,clip]{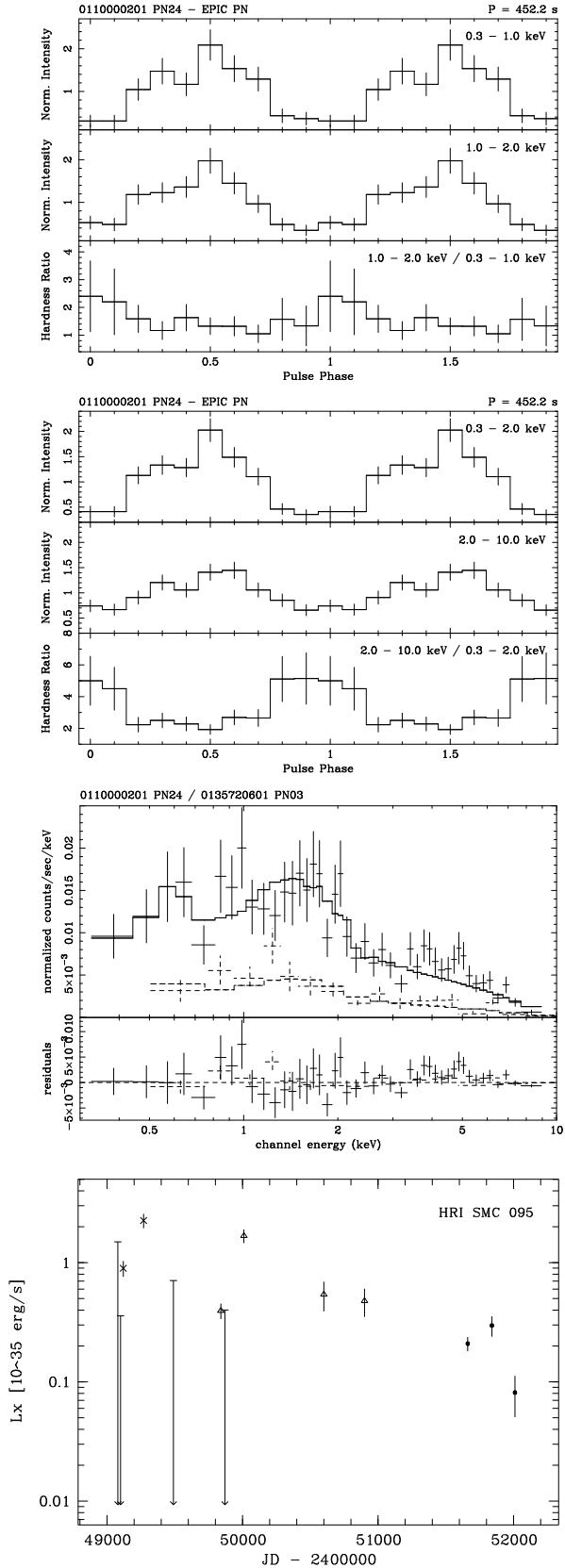}
\caption{\label{J0101-7211}
Folded light curves, spectra, and long term light curve (0.1 -- 2.4~keV) of 
RX\,J0101.3--7211 (source No 14).
Hardness ratio and symbols as in Fig\,\ref{J0054-7226}.
For the spectra, solid lines are used for the data
of the obs.\ ID 01100002, and dashed lines for obs.\ ID 01357206.
}
\end{figure}                                  

The ROSAT source \object{RX\,J0101.3--7211} is the first X-ray 
binary pulsar of which the discovery was based on XMM-Newton data. It was
covered in two additional observations finding the source again in a low
intensity state. The pulse
period of 455$\pm$2~s \citep{2001A&A...369L..29S} was  
verified in the new data of the observation ID 01100002: 452.2$\pm$0.5~s. 
During the observation
ID 01357206, the source was too faint for a timing analysis.
The folded light curves show strong variation in all bands 
(Fig.\,\ref{J0101-7211}). The spectrum of the source becomes harder during 
pulse minimum. The spectrum is well fitted with a soft thermal component 
described by a MEKAL model 
($kT = 0.20^{+0.09}_{-0.06}$~keV)
with a low metal abundance 
($0.11^{+0.10}_{-0.11}$ times solar)
and a power law component 
absorbed by a high column density (Table \ref{spectab}).  
The optical counterpart (OGLE\,01012064--7211187) is a Be-star. 

\subsection{Source No 15: RX\,J0101.6--7204}

\begin{figure}
\centering
\hspace{.7mm}\includegraphics[width=5.1cm,angle=270,clip]{h4167f12a.ps}\\*[2mm]
\hspace{-4.7mm}\includegraphics[width=5.65cm,angle=270,bb=40 50 560 772,clip]{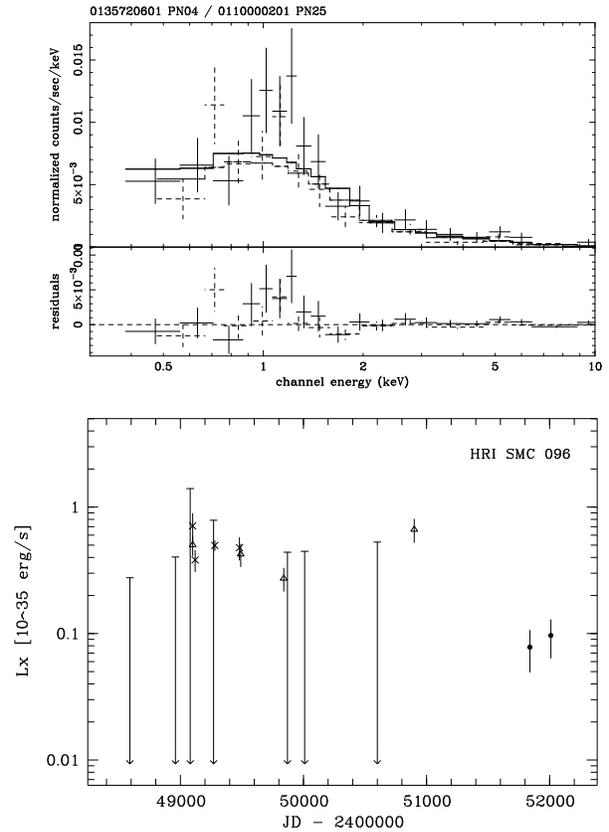}
\caption{\label{J0101-7204}    
Spectra and long term light curve (0.1 -- 2.4~keV) of RX\,J0101.6--7204 
(source No 15).
For the spectra, solid lines are used for the data
of the obs.\ ID 01357206, and dashed lines for obs.\ ID 01100002.
Symbols for the long term light curve as in Fig.\,\ref{J0054-7226}.
}
\end{figure}                                  

The Be/XRB candidate \object{RX\,J0101.6--7204} 
with an emission line star
at the ROSAT PSPC and HRI positions ([HS2000]), 
was observed
in two XMM-Newton pointings. Its spectrum and long
term light curve are shown in Fig.\,\ref{J0101-7204}. The spectrum can be
modelled as a moderately absorbed power law.
No pulsations were discovered. 

\subsection{Source No 16: AX\,J0103--722}  

\begin{figure}
\centering
\includegraphics[width=5.3cm,angle=270]{h4167f13a.ps}\\*[2mm]
\includegraphics[width=5.3cm,angle=270]{h4167f13b.ps}\\*[2mm]
\hspace{.7mm}\includegraphics[width=5.1cm,angle=270,clip]{h4167f13c.ps}\\*[2mm]
\hspace{-4.7mm}\includegraphics[width=5.65cm,angle=270,bb=40 50 560 772,clip]{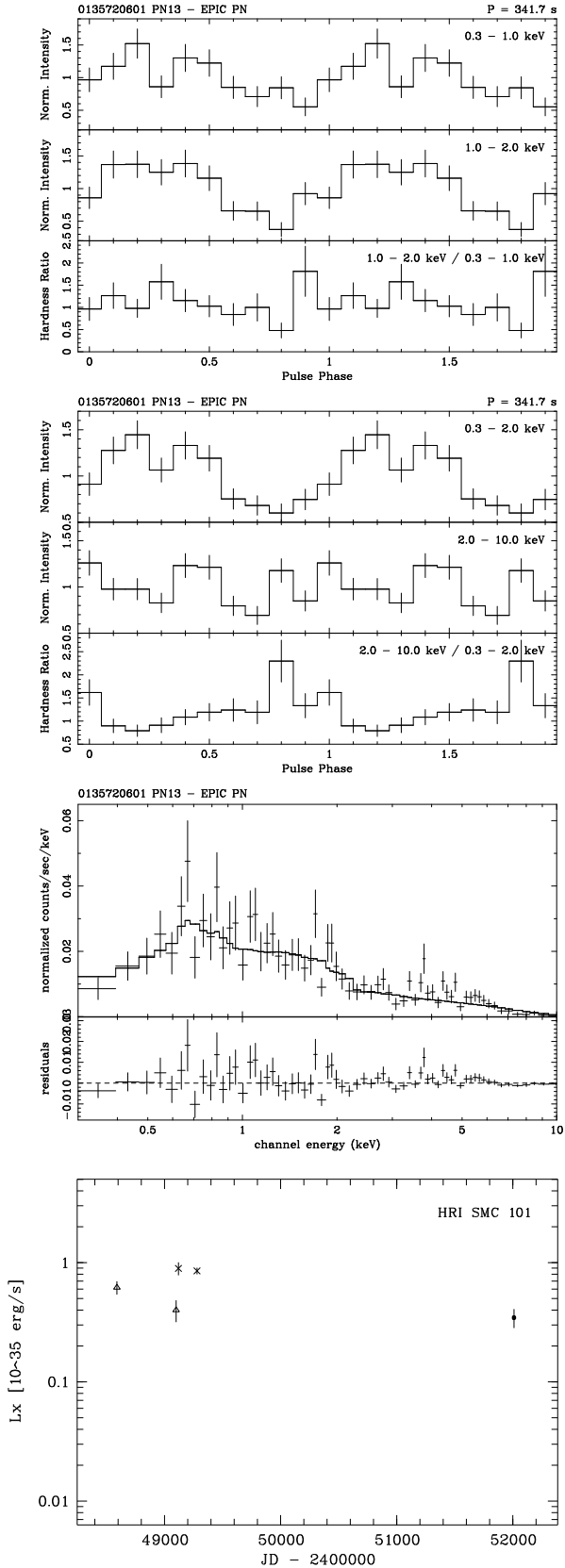}
\caption{\label{J0103-7209}
Folded light curves, spectrum, and long term light curve (0.1 -- 2.4~keV) of 
AX\,J0103--722 (source No 16). 
Hardness ratio and symbols as in Fig\,\ref{J0054-7226}.
}
\end{figure}                                  

For the Be/X-ray binary \object{AX\,J0103--722} 
a pulse period of 345.2$\pm$0.1~s was determined by 
\citet{1998IAUC.6999....1I}. In the XMM-Newton data, pulsations were confirmed 
with a period of 341.7$\pm$0.4~s. The folded light curves show strong 
variation below 2.0~keV (Fig.\,\ref{J0103-7209}), whereas
in the hard band, the variations are strongly reduced.
The spectrum is well reproduced with a power law and a thermal component
(see Table \ref{spectab}). 
The MEKAL model for the thermal component yields
$kT = 0.27_{-0.07}^{+0.08}$~keV and metal abundances of $0.31_{-0.17}^{+0.43}$ 
with respect to solar.

\subsection{Source No 17: RX\,J0103.6--7201}                 

\begin{figure}
\centering
\hspace{.7mm}\includegraphics[width=5.1cm,angle=270,clip]{h4167f14a.ps}\\*[2mm]
\hspace{-4.7mm}\includegraphics[width=5.65cm,angle=270,bb=40 50 560 772,clip]{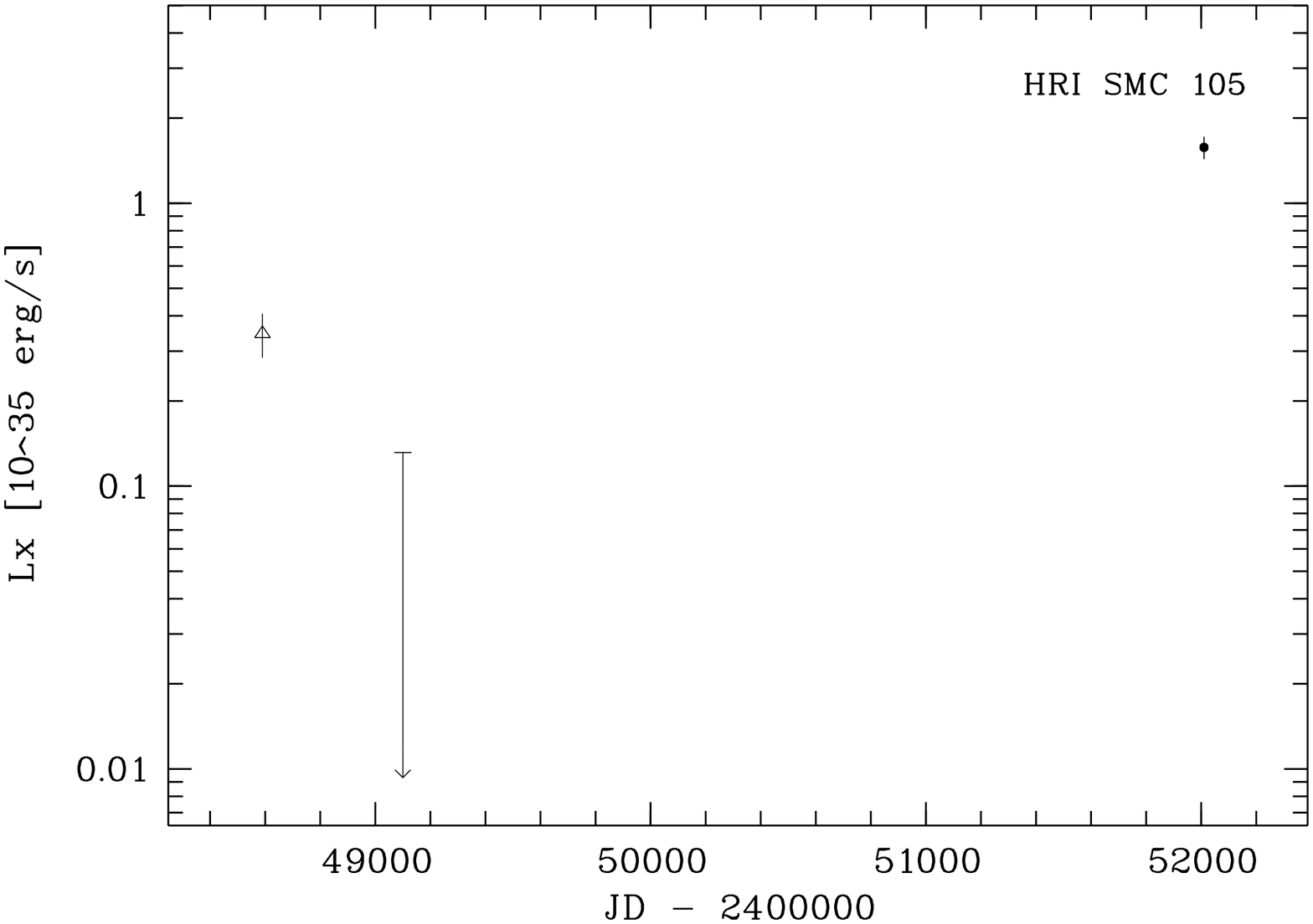}
\caption{\label{J0103-7201}    
Spectrum and long term light curve (0.1 -- 2.4~keV) of RX\,J0103.6--7201 
(source No 17).
Symbols for the long term light curve as in Fig.\,\ref{J0054-7226}.
}
\end{figure}                                  

For the HMXB candidate \object{RX\,J0103.6--7201} ([HS2000]),
an acceptable fit was obtained for the spectrum 
with a power law and a thermal component (Table \ref{spectab}). 
Modelling the thermal component with MEKAL, we obtained
$kT = 0.27_{-0.04}^{+0.03}$ and metal abundances of $0.77_{-0.28}^{+0.17}$
times solar. Since the source was bright, we also used the VMEKAL model 
instead of the MEKAL model, which allows to determine the
abundance for each of the elements. This resulted 
in an improvement of the fit, the model reproducing the peaks   
around 0.6 and 0.9~keV. 
The photon indices $\Gamma$ and the absorbing column densities $N_{\rm H}$ 
for both fits are comparable, as can be seen in Table \ref{spectab}. 
Also the temperature values $kT$ agree well for 
MEKAL and VMEKAL within the 1~$\sigma$ errors. 
The spectrum with the power law + VMEKAL fit is shown
in Fig.\,\ref{J0103-7201}.
The comparison to ROSAT data shows that this source was in high 
luminosity state during the XMM-Newton observation with 
$L_{\rm X} = 1.1 \times 10^{36}$~erg~s$^{-1}$ (0.3 -- 10.0~keV). In spite of 
the high photon statistics with 3,300 counts, no pulsations were discovered.  
Also the analysis of the events separated into soft, medium, and hard band
revealed no pulsations.

\subsection{Source No 18: RX\,J0105.9--7203}                 

\object{RX\,J0105.9--7203} is a HMXB candidate, coinciding with an emission 
line object. Since the source was 
very faint during the XMM-Newton observation, the photon statistics are very 
low and no timing analysis was possible. 
The PSPC count rate derived from the ROSAT observation was 
$4.01 \times 10^{-3} \pm 0.56 \times 10^{-3}$~s$^{-1}$ 
\citep{2000A&AS..142...41H}, corresponding to 
$\sim5 \times 10^{-2}$~s$^{-1}$ for XMM-Newton EPIC PN 
(thin1 filter). With a count rate
of $3.05 \times 10^{-2} \pm 0.28 \times 10^{-2}$~s$^{-1}$ 
(Table \ref{xbtab}), the source was fainter 
during the XMM-Newton observation.

\section{Sources No 6 and 12: New HMXB candidates}

\begin{figure}
\centering
\hspace{.7mm}\includegraphics[width=5.2cm,angle=270,clip]{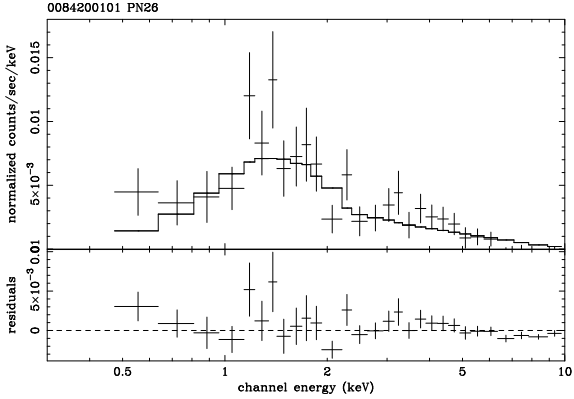}
\caption{\label{J0057-7219}
Spectrum of XMMU\,J005735.7--721932 (source No 6).
}
\end{figure}                                  

To identify a HMXB, it is crucial to find an optical counterpart and confirm
that it is an early-type star. If an emission line object is found at the
position of a hard X-ray source and other objects are ruled out as counterpart,
the source is presumably a Be/XRB. 
Cross-correlating the XMM-Newton source list with the emission line star 
catalogue of \citet{1993A&AS..102..451M}, we discovered two new sources which 
met these criteria. 

\object{XMMU\,J005735.7--721932} (source No 6) 
is found at the position of [MA93]1020 and 
likely coincides with the source No 19 in \citet{2000ApJS..128..491Y}. 
\object{XMMU\,J010030.2--722035} (source No 12) is associated with [MA93]1208. 
Both X-ray sources are very faint. 
Only for XMMU\,J005735.7--721932, we had enough
counts to extract a spectrum (Fig.\,\ref{J0057-7219}).
The best fit model is a moderately absorbed power law (Table \ref{spectab}).
Furthermore, Chandra data showed that this source has pulsed
emission with a period of 564.81$\pm$0.41~s \citep{2003ApJ...584L..79M}.
This period could not be confirmed in the XMM-Newton data.

As indicated by the hardness ratio HR1, these sources have a hard spectrum.
Therefore, these two sources are suggested as new Be/XRB candidates.
For further investigation, we need to perform follow-up
observation in the optical band in order to verify if the emission line
objects are Be stars. 

\section{Discussion}

\begin{figure}
\centering
\hspace{-2.2mm}\includegraphics[width=6.1cm,angle=270,clip]{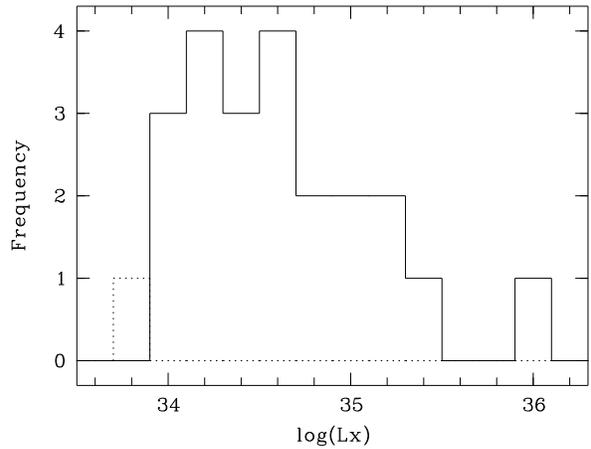}
\caption{\label{lxhisto}
Histogram of luminosities [erg~s$^{-1}$] of the HMXBs and candidates in the 
XMM-Newton band (0.3 -- 10.0~keV). The upper limit for SMC\,X-2 is shown with
dashed line.
}
\end{figure}

The comparison between the XMM-Newton sources detected with ML $>$ 10 and other
X-ray catalogues 
\citep{1992ApJS...78..391W,1999A&AS..136...81K,2000A&AS..142...41H,2000A&AS..147...75S}
demonstrates that we detected all known HMXBs and candidates which exist in the
four observed fields, except for SMC\,X-2 which was very faint. 
SMC\,X-2 was marginally detected with a likelihood of ML = 3.4, and we 
derived an upper limit of $6.5 \times 10^{33}$~erg~s$^{-1}$ (0.3 -- 10.0~keV). 
The luminosities of all the other sources in the 0.3 -- 10.0~keV 
band are higher than $8 \times 10^{33}$~erg~s$^{-1}$ 
at an assumed distance
of 60~kpc \citep{1999IAUS..190..569V}, 
as is shown in 
Fig.\,\ref{lxhisto}. As we have seen in the long term light curves, most
of the sources were in quiescence during the XMM-Newton observations,
whereas they were mostly detected during outburst by previous missions.
This indicates that all known HMXBs in the SMC have luminosities higher than  
$\sim7 \times 10^{33}$~erg~s$^{-1}$ in quiescence and can be detected by  
XMM-Newton in observations with an exposure of about 15~ks.
Consequently, we have an extensive set of HMXBs for studying their 
properties.

In order to visualise the spectral characteristics of the HMXBs, we plotted
the hardness ratios HR1, HR2, and HR3 in Fig.\,\ref{hr1hr2hr3}. 
The high absorption in XRBs causes positive values for HR1,
while HR2 and HR3 have small absolute values around zero.
AGNs typically show steeper X-ray spectra than HMXBs. Therefore, the 
two source classes can be disentangled using hardness ratios. This can 
also be applied to classification work on other nearby galaxies.

\subsection{Pulsations and soft energy emission}

X-ray spectra of high mass X-ray binaries below 10~keV can be in general 
modelled as a 
power law with a photon index of $\Gamma = 0 - 2$. For HMXBs located far away  
from the Galactic plane or in the Magellanic Clouds, the interstellar  
absorption in the line of sight is low, and an additional soft spectral 
component was discovered in supergiant X-ray binary systems like
SMC\,X-1 \citep{1983ApJ...266..814M,1995ApJ...445..896W} or LMC\,X-4 
\citep{1996ApJ...467..811W}, as well as in Be/X-ray binary systems like
RX\,J0059.2--7138 \citep{2000PASJ...52..299K} or 
EXO\,053109--6609.2 \citep{2003A&A...........H}.

In our sample of HMXBs and candidates in the SMC, pulsations were confirmed for
six sources. Studying the pulsations and hardness ratio changes in different
bands, we found that there are different types of pulsations. Furthermore,
four out of these six were bright enough to allow us to test
the existence of a soft component in their spectra. The pulsating sources
of our sample can be divided into four groups: 
\begin{enumerate}
\item There are pulsations in all bands and the ratios between the 
harder and softer bands are almost constant (RX\,J0054.9--7226, 
Fig.\,\ref{J0054-7226} and XMMU\,J005605.2--722200, 
Fig.\,\ref{J0056-7222}). Because of low photon statistics, spectral 
analysis of these sources yielded no significant results. 
\item Pulsations are discovered below 2~keV (AX\,J0058--720, 
Fig.\,\ref{J0058-7202} 
and AX\,J0103--722, Fig.\,\ref{J0103-7209}). There might 
be pulsations also in the hardest band, although not significant due to low
statistics. The hardness ratios seem to become higher during pulse 
minimum. The spectra of these sources include a thermal component.
If it is confirmed that pulsations are in fact existent only in the soft band,
this will suggest that their origin is of small size,
probably locally illuminated surface or surroundings of the neutron star.
\item In the case of RX\,J0057.8--7207 
(Fig.\,\ref{J0057-7207}), there are flux variations in all bands. Most  
pronounced pulsations are found above 2~keV with correlated increase of the 
hardness ratio, which follows the pulse shape of the hard band.
Since the spectrum is a perfect power law, the coincidence of the maximum of
the hardness ratio and that of the pulse might indicate a variation of the
absorption ($N_{\rm H}$).
\item RX\,J0101.3--7211 (Fig.\,\ref{J0101-7211}) shows 
clear pulsations in all bands and becomes harder significantly at pulse
minimum. In its spectrum, there is a low energy excess besides the power law
component, which can be modelled as a thermal emission. The pulsations in the
soft band might be caused by $N_{\rm H}$ variations as well as by changes in
the soft emission component.
\end{enumerate}
The low energy component, 
which seems to be thermal was also found in the spectrum
of RX\,J0103.6--7201. This source is variable on long timescales,
as can be seen in Fig.\,\ref{J0103-7201}. However, pulsations on short 
timescales were not discovered, although the source was brighter during the
XMM-Newton observation than in former observations. 

\subsection{Origin of the soft emission}

The low energy component in the spectrum of the supergiant systems SMC\,X-1 
\citep{1983ApJ...266..814M} and LMC\,X-4 
\citep{1996ApJ...467..811W} was modelled as blackbody emission or 
thermal Bremsstrahlung which arises from the stellar wind of the supergiant,
the accretion disk, or the fan-beam of the 
accretion column close to the neutron star surface. 
However, \citet{2002ApJ...579..411P} pointed out that a power law 
nature is most probable for the soft emission. 
They derived that the pulse shape of the soft
emission from SMC\,X-1 is sinusoidal, similar to the soft energy light 
curve of Her\,X-1 \citep[e.g.][]{1997A&A...327..215O}.
In our Galaxy, the supergiant system Vela\,X-1 is thought to show emission 
from the atmosphere and stellar wind of the companion as well as from the gas 
stream towards the neutron star 
\citep[][and references therein]{1990ApJ...361..225H}.
High resolution spectroscopy of Galactic HMXBs like 
Cen\,X-3 \citep[with Chandra HETG,][]{2002APS..APRN17069W} 
or Her\,X-1 \citep[with XMM-Newton RGS,][]{2002ApJ...578..391J} 
resolved fluorescent lines and hydrogen- and helium-like lines of elements 
from Ne to Fe.
The line fluxes of Cen\,X-3 are consistent with recombination 
radiation from photo-ionised and collisionally ionised plasma as well as 
resonant line scattering in photo-ionised plasma \citep{2002APS..APRN17069W}. 

As for the Be/X-ray binary systems, \citet{2000PASJ...52..299K} analysed both
ASCA and ROSAT data of RX\,J0059.2--7138 and found that there is a soft 
component in the spectrum, which can be modelled as a thermal emission with
$kT = 0.37$~keV. Below 2.0~keV, the source shows no pulsations. 
Therefore they argue
that the soft emission originates from a large region comparable to the full 
binary system. Using an XMM-Newton observation of a northern field in the LMC, 
\citet{2003A&A...........H} extracted emission from the Be/X-ray binary 
EXO\,053109--6609.2 and showed that there are strong pulsations above 0.4~keV. 
In the spectrum there is a low energy thermal component, which is believed to 
arise from the equatorial disk around the Be star,
illuminated by the X-ray source. 

The origin of the soft emission from HMXBs is not clearly understood. One would
expect that there are differences between a supergiant and a Be system.
Most of the HMXBs which have been studied in detail (since they
are located in the Milky Way and therefore closer) are supergiant systems,
whereas the sources in the SMC we are confronted with, are Be systems.
In Be/XRBs, the neutron star and the Be star are thought to form a binary   
system with an extended orbit. This makes the stellar material in the 
equatorial disk around the Be star as the origin of the soft pulsed emission 
rather implausible. The HMXBs in the MCs are ideal objects to study the soft 
part of their spectrum, since the absorption by Galactic foreground matter is 
low in the direction of the MCs.
The existence of a soft thermal component in the spectrum and pulsations
below 1 -- 2~keV in our data indicates that the size of the origin of the 
soft emission  
is not as large as is assumed for e.g.\ RX\,J0059.2--7138. In addition to 
timescales and luminosities, a crucial parameter for the 
physical processes responsible for this emission is the magnetic field of the 
neutron star.
In order to clarify the conditions in which the soft component is produced, 
at least we need to get information about the orbital motion and about a  
possible orbital phase dependence of the total source spectrum as well as the  
pulsed emission. As for the SMC Be systems discussed here, the orbital period 
is known only for one source.

\subsection{OB systems vs.\ Be systems}

In the last few years, the number of known Be/XRBs in the SMC increased 
drastically based on temporal studies of hard X-ray sources and optical
observations. In order to identify an X-ray source
as a HMXB and clarify the nature of the mass donor star, we need to perform
spectroscopy of the optical counterpart. Since most of the HMXB candidates 
which are known now are correlated to emission line objects, we expect that  
additional Be/XRBs will be found in the near future. 
This will further increase the ratio between the Be systems and the 
OB systems among the HMXBs in the SMC. 
Be/XRBs are thought to evolve from binary systems in about
$1.5 \times 10^{7}$~yrs, whereas supergiant systems 
evolve faster due to the high mass of the companion star. The large number
of Be/XRBs sets constraints on the secondary star formation in the SMC,   
making a burst some $10^{7}$~yrs ago most likely.

\section{Summary}

We analysed XMM-Newton EPIC PN and MOS 1/2 data of four pointings towards the  
SMC. One observation covered the field around the HMXB SMC\,X-2 in the
south, whereas the fields of view of the other three are located in the 
northern 
part of the main body of the SMC. In total, there were 15 detections which 
were identified as known HMXBs or XRB candidates. For SMC\,X-2 which 
was faint during the observation, a flux upper limit of  
$1.5 \times 10^{-14}$~erg~cm$^{-2}$~s$^{-1}$ (0.3 -- 10.0~keV) was derived. 
We found two new sources (XMMU\,J005735.7--721932 and 
XMMU\,J010030.2--722035) which have a hard spectrum and positionally  
coincide with emission line objects \citep{1993A&AS..102..451M}. These sources
are proposed as new HMXB candidates, probably Be systems. 

Four sources in our list were known to show pulsed emission and pulse periods
had been determined in former observations. In this work, the pulse periods 
were confirmed for all four sources. Furthermore, we discovered that two
other sources which had been proposed to be Be/XRB candidates, show pulsations:
XMMU\,J005605.2--722200 with a pulse period of 140.1$\pm$0.3~s and 
RX\,J0057.8--7207 with 152.34$\pm$0.05~s.
 
Spectral analysis of the sources was performed. For faint sources,
a good fit was obtained with a single power law spectrum. However, for three 
brighter sources, we could show that there is a significant low energy excess
in the spectrum, if we only assume a power law. The spectra indicate emission 
line features,
suggesting that the emission is thermal. This soft component was modelled
as thermal emission, yielding temperatures of 0.2 -- 0.3~keV. The abundances
in the emitting plasma are below solar values, but comparable to typical SMC  
values \citep{1992ApJ...384..508R}: for RX\,J0101.3--7211 
it is $0.11^{+0.10}_{-0.11}$ times solar, and for AX\,J0103--722  
best fit is obtained with $0.31_{-0.17}^{+0.43}$ times solar. 
The errors are 1~$\sigma$ values.
Only for 
RX\,J0103.6--7201 the abundance
is higher with $0.77_{-0.28}^{+0.17}$ with respect to solar. 

The flux of the sources in the MCs is low compared to the bright 
($L_{\rm X} = 10^{37-38}$~erg~s$^{-1}$) HMXBs in 
our Galaxy, making it difficult to 
perform a detailed analysis of their soft emission. However, 
the sources in the MCs have the advantage of low Galactic
absorption. This allows us to study the thermal emission from a large 
sample of HMXBs and to increase the understanding of the interaction
between X-rays from the compact object and the ambient stellar matter. It is 
also important to verify if there is a change in temperature or emissivity,
which is related to the orbital phase of the binary system.
Due to the improved time resolution and sensitivity, there is a large 
detection potential for new pulsating XRBs in further XMM-Newton observations.

\begin{acknowledgements}
We would like to thank the anonymous referee for useful comments.
The XMM-Newton project is supported by the Bundesministerium f\"ur
Bildung und Forschung / Deutsches Zentrum f\"ur Luft- und Raumfahrt
(BMBF/DLR), the Max-Planck Society and the Heidenhain-Stiftung.   
This research has been carried out by
making extensive use of the SIMBAD data base operated at CDS,
Strasbourg, France.
The Digitized Sky Survey was produced at the Space Telescope Science 
Institute under U.S. Government grant NAG W-2166. The images of these surveys 
are based on photographic data obtained using the Oschin Schmidt Telescope on 
Palomar Mountain and the UK Schmidt Telescope. The plates were processed into 
the present compressed digital form with the permission of these institutions.
This research has made use of data obtained through the High Energy 
Astrophysics Science Archive Research Center Online Service, provided by the
NASA/Goddard Space Flight Center. 
\end{acknowledgements}

%\bibliography{/home/manami/tex/bibtex/my,/home/manami/tex/bibtex/xb,/home/manami/tex/bibtex/lmchricat2000,/home/manami/tex/bibtex/smchricat2000}
%\bibliography{/home/msasaki/MPEds02/tex/bibtex/my,/home/msasaki/MPEds02/tex/bibtex/xb,/home/msasaki/MPEds02/tex/bibtex/lmchricat2000,/home/msasaki/MPEds02/tex/bibtex/smchricat2000}

\end{document}